\documentclass[twocolumn]{aastex63}

\usepackage{xcolor, fontawesome}
\definecolor{twitterblue}{RGB}{64,153,255}

\newcommand{\twitter}[1]{\href{https://twitter.com/#1 }{\textcolor{twitterblue}{\faTwitter}\,\tt \textcolor{twitterblue}{@#1}}}

\newcommand{\github}[1]{\href{https://github.com/#1 }{\textcolor{black}{\faGithub}\,\tt \textcolor{black}{@#1}}}

\newcommand\degrees[1]{\ensuremath{#1^\circ}}
\newcommand{\as}{$^{\prime\prime}$}

\submitjournal{AAS Journals}
\accepted{18 Sept. 2020}

\shorttitle{ALMA Survey of $\lambda$ Orionis Disks}
\shortauthors{Ansdell et al.}

\begin{document}

\title{An ALMA survey of $\lambda$ Orionis disks: from supernovae to planet formation}

\correspondingauthor{Megan Ansdell; \\
\twitter{megaparsec808}; \github{mansdell}}
\email{megan.c.ansdell@nasa.gov}

\author[0000-0003-4142-9842]{Megan Ansdell}
\affiliation{Flatiron Institute, Simons Foundation, 162 Fifth Ave, New York, NY 10010, USA}
\affiliation{NASA Headquarters, 300 E Street SW, Washington, DC 20546, USA}

\author{Thomas J. Haworth}
\affiliation{Astronomy Unit, School of Physics and Astronomy, Queen Mary University of London, London E1 4NS, United Kingdom}

\author{Jonathan P. Williams}
\affiliation{Institute for Astronomy, University of Hawai`i at M\={a}noa, Honolulu, HI, USA}

\author{Stefano Facchini}
\affiliation{European Southern Observatory, Karl-Schwarzschild-Str. 2, D-85748 Garching bei M\"{u}nchen, Germany}

\author{Andrew Winter}
\affiliation{Astronomisches Rechen-Institut, Zentrum f\"{u}r Astronomie der Universit\"{a}t Heidelberg, M\"{o}nchhofstra\ss e 12-14, 69120 Heidelberg, Germany}

\author{Carlo F. Manara}
\affiliation{European Southern Observatory, Karl-Schwarzschild-Str. 2, D-85748 Garching bei M\"{u}nchen, Germany}

\author{Alvaro Hacar}
\affiliation{Leiden Observatory, Leiden University, P.O. Box 9513, 2300-RA Leiden, The Netherlands}

\author{Eugene Chiang}
\affiliation{Department of Astronomy, University of California at Berkeley, Berkeley, CA 94720, USA}

\author{Sierk van Terwisga}
\affiliation{Max-Planck-Institute for Astronomy, K\"{o}nigstuhl 17, 69117 Heidelberg, Germany}

\author{Nienke van der Marel}
\affiliation{Department of Physics \& Astronomy, University of Victoria, Victoria, BC, V8P 1A1, Canada}

\author{Ewine F. van Dishoeck}
\affiliation{Leiden Observatory, Leiden University, P.O. Box 9513, 2300-RA Leiden, The Netherlands}


\begin{abstract}

Protoplanetary disk surveys by the Atacama Large Millimeter/sub-millimeter Array (ALMA) are now probing a range of environmental conditions, from low-mass star-forming regions like Lupus to massive OB clusters like $\sigma$~Orionis. Here we conduct an ALMA survey of protoplanetary disks in $\lambda$~Orionis, a $\sim$5~Myr old OB cluster in Orion, with dust mass sensitivities comparable to the surveys of nearby regions ($\sim$0.4~$M_\oplus$). We assess how massive OB stars impact planet formation, in particular from the supernova that may have occurred $\sim$1~Myr ago in the core of $\lambda$~Orionis; studying these effects is important as most planetary systems, including our Solar System, are likely born in cluster environments. We find that the effects of massive stars, in the form of pre-supernova feedback and/or a supernova itself, do not appear to significantly reduce the available planet-forming material otherwise expected at the evolved age of $\lambda$~Orionis. We also compare a lingering massive ``outlier" disk in $\lambda$~Orionis to similar systems in other evolved regions, hypothesizing that these outliers host companions in their inner disks that suppress disk dispersal to extend the lifetimes of their outer primordial disks. We conclude with numerous avenues for future work, highlighting how $\lambda$~Orionis still has much to teach us about perhaps one of the most common types of planet-forming environments in the Galaxy.
\end{abstract}



\section{Introduction} 
\label{sec:intro}

Thousands of diverse exoplanetary systems have now been discovered \citep[e.g., see review in ][]{WF2015}, yet how they all formed remains unclear due to our still-incomplete understanding of the evolution of the progenitor protoplanetary disks. These disks are traditionally thought to evolve through viscous accretion \citep[e.g.,][]{LBP1974}, where turbulence redistributes angular momentum and drives material onto the central star. However other processes, both internal and external to the disks, can also significantly influence their evolution. In particular, the external influences of massive OB stars on the disks of surrounding lower-mass stars is important to study as many planetary systems, including our Solar System, are likely born in cluster environments \cite[e.g.,][]{Adams2010,Winter2020}. 

In stellar clusters, the ultraviolet (UV) emission from the OB stars induces thermal winds from nearby disks, which effectively remove planet-forming material from their outer regions in a process called ``external photoevaporation" \cite[e.g.,][]{Hollenbach1994}. Theoretical work suggests that external photoevaporation can severely shorten disk lifetimes and truncate outer disk radii in cluster environments, whereas stellar encounters play a relatively insignificant role in sculpting disk populations \citep[e.g.,][]{SC2001,CM2019}, even in moderate UV environments \cite[e.g.,][]{Facchini2016,Haworth2018,Winter2018}. Moreover, external photoevaporation of the outer disk is theorized to dominate over viscous spreading under realistic cluster conditions and dust grain growth prescriptions \citep[e.g.,][]{Clarke2007,Facchini2016,Winter2018}. While direct detection of photoevaporative disk winds is observationally challenging, a handful of cases exist \citep[e.g.,][]{Henney1999, Rigliaco2009}. Meanwhile, indirect evidence of external photoevaporation from observations of its expected impact on other disk properties is growing \citep[e.g.,][]{Mann2014, Kim2016,Guarcello2016,Ansdell2017,vanTerwisga2019}.
 
Still, our understanding of how disk evolution is altered when one of these rapidly evolving OB stars inevitably goes supernova remains limited, due in part to the lack of recent supernova events in nearby star-forming regions (SFRs) available for study. This makes the $\lambda$~Orionis cluster a key target, as a supernova is thought to have occurred in the core of the region $\sim$1~Myr ago \citep[e.g.,][]{CS1996,DM2001}. At $\sim$5~Myr of age \cite[e.g.,][]{Hernandez2009}, $\lambda$~Orionis may therefore provide a rare snapshot of an evolved SFR post-supernova, with the remaining OB stars, lower-mass stellar population, and remnant molecular cloud all still present. Supernovae are theorized to strip significant amounts of mass from disks around nearby stars via ram pressure \citep[e.g.,][]{CP2017} as well as expose them to enhanced cosmic ray ionization rates that accelerate carbon processing of CO into other molecules \citep[e.g.,][]{Eistrup2016,Scharz2018,Bosman2018}, thereby providing predictions to test against observations.

A partial survey of the disk population in $\lambda$~Orionis was conducted at millimeter wavelengths by \cite{Ansdell2015} with JCMT/SCUBA-2. Observations of disks at these longer wavelengths are particularly useful because any optically thin continuum emission can be related to the available planet-forming solids in the disk \cite[e.g.,][]{Hildebrand1983, Andrews2015}, while various molecular lines can probe gas content and/or chemistry \cite[e.g.,][]{Miotello2017, vanTerwisga2019, Miotello2019, Booth2020}. However, due to the evolved age ($\sim$5~Myr) and large distance ($\sim$400~pc) of $\lambda$~Orionis, combined with the limited sensitivity of JCMT/SCUBA-2, \cite{Ansdell2015} detected only one disk in their survey. Fortunately,  the Atacama Large Millimeter/sub-millimeter Array (ALMA) now provides the high sensitivity required to efficiently survey the disk population in $\lambda$~Orionis to dust mass sensitivities that are commonly achieved for the nearby ($\sim$150~pc) SFRs like Lupus \citep[][]{Ansdell2016,Ansdell2018}, Chamaeleon~I \citep{Pascucci2016}, $\rho$~Ophiuchus \citep{Cieza2019}, and Upper Sco \citep{Barenfeld2016}. This paper presents the results of such an ALMA survey. 

We begin in Section~\ref{sec:history} by discussing the formation and evolution of $\lambda$~Orionis, as the cluster's history is central to our analysis. In Section~\ref{sec:sample}, we describe our sample of protoplanetary disks and their host star properties. Our ALMA survey of these disks and the key observational results are presented in Section~\ref{sec:alma}, while the basic disk properties are derived in Section~\ref{sec:properties}. We then examine the implications for our understanding of how OB stars affect disk evolution in Section~\ref{sec:discussion}, which also discusses how massive ``outlier" disks in evolved regions like $\lambda$~Orionis may improve our knowledge on the pathways of planet formation and disk dispersal. We summarize our findings and provide avenues for future work in Section~\ref{sec:summary}.


\section{The $\lambda$~Orionis Cluster} \label{sec:history}

\begin{figure*}
\begin{center}
\includegraphics[width=18cm]{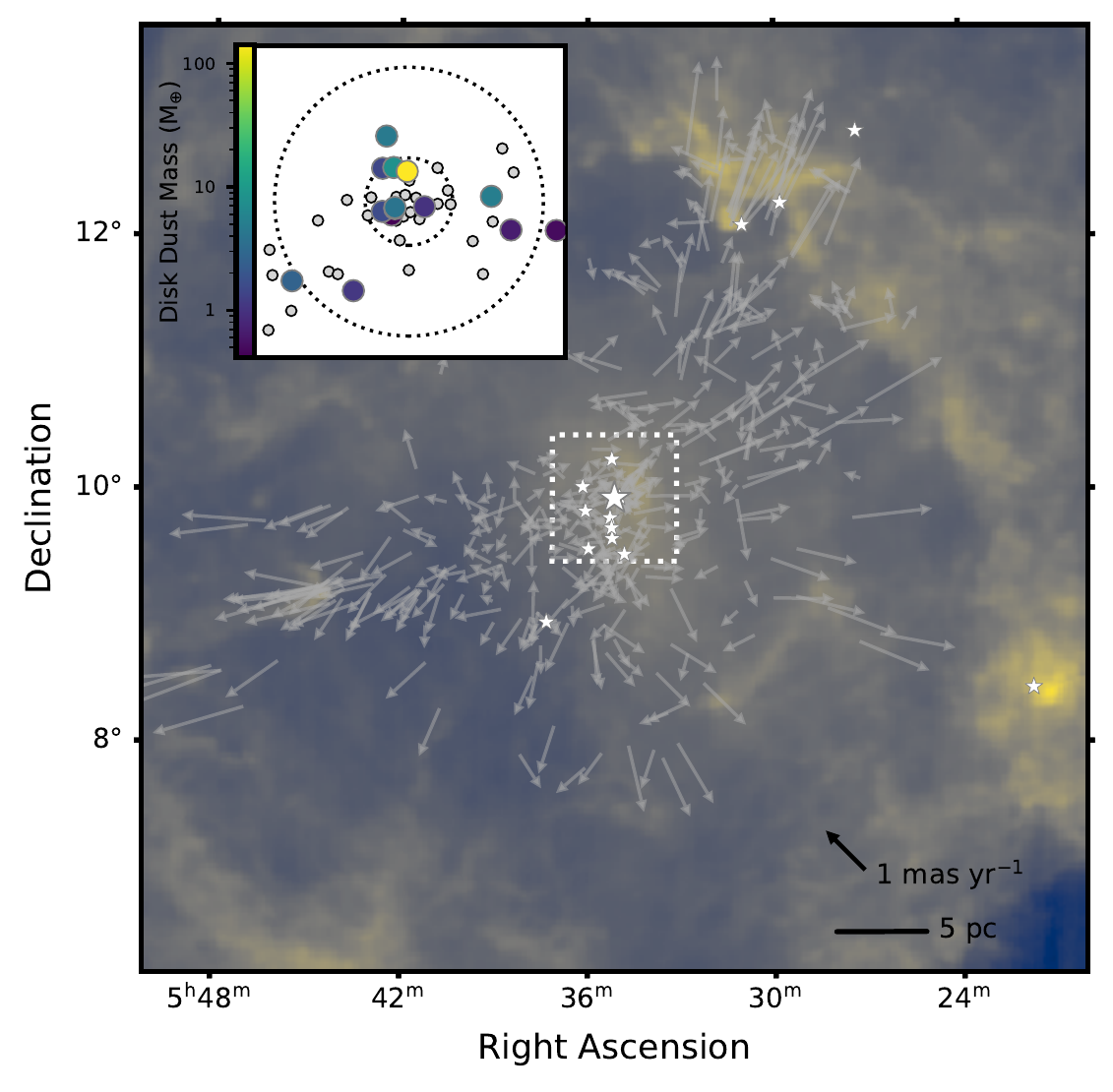}
\caption{An IRAS 100-$\mu$m image of the $\lambda$~Orionis cluster. The white dashed box outlines the region surveyed for disks with {\it Spitzer} in \cite{Hernandez2009,Hernandez2010}, which we use as the basis of our ALMA survey sample selection (Section~\ref{sec:disks}). The inset zooms in on this region: the large circles are our ALMA detections color-coded by dust mass (Section~\ref{sec:mass}), while the small gray circles are the non-detections (Section~\ref{sec:cont}); the dotted lines represent 1~pc and 3~pc radial distances from the $\lambda$~Ori system at the center of the cluster. Gray arrows show the {\it Gaia} DR2 proper motions of the cluster members identified by \cite{Kounkel2018} in the LSR frame with the median cluster value subtracted; a reference vector of magnitude 1~mas~yr$^{-1}$ and a scale bar of 5~pc are shown in the lower right corner. The small white stars are the locations of the B-type stars in the region, and the larger white star is the location of the $\lambda$~Ori system, which contains the only O-type star in the region (Section~\ref{sec:history}).}
\label{fig:pm}
\end{center}
\end{figure*}

The formation, evolution, and current state of the $\lambda$~Orionis cluster has been studied and debated in many works \citep[e.g.,][]{MM1987,CS1996,DM2001,Hernandez2009,Bayo2011,Mathieu2015,Kounkel2018}. In this section, we briefly summarize the existing observations of the cluster, then assume a commonly adopted but still-debated interpretation, which has implications for our later analysis.

As illustrated in Figure~\ref{fig:pm}, $\lambda$~Orionis currently consists of several hundred low-mass members centered on a core of OB stars. The most massive is the O8III star, $\lambda$~Ori, the brightest in the head of the Orion constellation. Although $\lambda$~Ori is the only O-type star in the cluster, it has a B0V companion at 1900~au projected separation, and this binary system is accompanied by nine other B-type stars (some close binaries themselves) in a dense $\sim$2~pc radius clump in the cluster center \citep{MP1977}. The interstellar dust and molecular gas in the region is largely confined to a $\sim$30~pc radius ring \citep[or possibly shell;][]{Lee2015}, which is centered on the clump of OB stars, encompasses the lower-mass population, and is rapidly expanding at $\sim$14~km~s$^{-1}$ \citep[e.g.,][]{MM1987,Lang2000}. As shown in Figure~\ref{fig:pm}, many of the $\lambda$~Orionis members also have proper motions directed radially away from the cluster center, with the more distant stars moving away the fastest \citep{Kounkel2018}.

The age of the lower-mass stellar population in $\lambda$~Orionis is $\sim$5~Myr \citep[e.g.,][]{DM2001,Hernandez2009}. This is largely based on Str\"{o}mgren photometry of the most massive OB stars, which suggests their formation occurred $\sim$6~Myr ago, combined with photometric and spectroscopic surveys of the surrounding pre-main sequence stars, which indicate an epoch of lower-mass star formation that began soon after but was halted $\sim$1--2~Myr ago \citep{DM1999,DM2001}. In this work we do {\em not} consider the younger dark Barnard~30 and Barnard~35 clouds, located along the edge of the ring, as part of the $\lambda$~Orionis cluster. 

We adopt the common interpretation in the literature of these observations, which is that an O-type star exploded as a supernova $\sim$1~Myr ago in the cluster core, carving out the central cloud to create the molecular ring/shell, while terminating star formation in the region \citep{CS1996,DM1999,DM2001,Mathieu2008,Mathieu2015}. However, we note that pre-supernova feedback may also carve out central cavities in molecular clouds \citep[e.g.,][]{Dale2014}. Moreover, although the radial motions of the $\lambda$~Orionis members have been attributed to a ``single-trigger" expansion caused by the supernova \citep{Kounkel2018}, such kinematics can result from any mechanism that removes the interstellar gas and its gravitational potential \citep{Winter2019}, and outward acceleration may even be aided by gravitational feedback of the dispersed gas \citep{Zamora2019}. While these and other caveats are discussed further in Section~\ref{sec:sn}, the  massive stars in $\lambda$~Orionis, whether through pre-supernova feedback and/or a supernova event itself, have played a key role in shaping the region.

\begin{deluxetable*}{lcccccccrr} 
\tabletypesize{\footnotesize} 
\centering 
\tablewidth{500pt} 
\tablecaption{Source Properties \label{tab-main}} 
\tablecolumns{10}  
\tablehead{ 
 \colhead{Source} 
&\colhead{2MASS ID} 
&\colhead{R.A.} 
&\colhead{Decl.} 
&\colhead{SpT} 
&\colhead{$M_{\star}$} 
&\colhead{$v_{\rm LSR}$} 
&\colhead{Ref.$^\dagger$} 
&\colhead{$\mu_\alpha$} 
&\colhead{$\mu_\delta$} 
\\ 
 \colhead{} 
&\colhead{} 
&\colhead{} 
&\colhead{} 
&\colhead{} 
&\colhead{($M_\odot$)} 
&\colhead{(km~s$^{-1}$)} 
&\colhead{} 
&\colhead{(mas yr$^{-1}$)} 
&\colhead{(mas yr$^{-1}$)} 
} 
\startdata 
LO 65 & J05331515+0950301 & 05:33:15.15 & +09:50:30.05 & M4 & 0.19 & 11.5 $\pm$ 0.2 & 2,4 & -0.64 $\pm$ 0.28 & -1.00 $\pm$ 0.23 \\
LO 1079 & J05334791+1001396 & 05:33:47.91 & +10:01:39.69 & M5.5 & 0.07 & ... & 1 & 2.11 $\pm$ 0.49 & -2.68 $\pm$ 0.41 \\
LO 1152 & J05334992+0950367 & 05:33:49.93 & +09:50:36.79 & M3.0 & 0.29 & 7.5 $\pm$ 0.1 & 1,4 & 2.65 $\pm$ 0.10 & -3.31 $\pm$ 0.08 \\
LO 1359 & J05335661+1006149 & 05:33:56.61 & +10:06:14.91 & M5.25 & 0.08 & ... & 1 & -0.68 $\pm$ 1.16 & -1.90 $\pm$ 0.83 \\
LO 1589 & J05340393+0952122 & 05:34:03.94 & +09:52:12.30 & M2.5 & 0.35 & 8.2 $\pm$ 0.3 & 1,4 & 2.82 $\pm$ 0.12 & -3.25 $\pm$ 0.09 \\
LO 1624 & J05340495+0957038 & 05:34:04.96 & +09:57:03.76 & M3 & 0.29 & 12.0 $\pm$ 0.1 & 2,4 & 0.27 $\pm$ 0.07 & -2.14 $\pm$ 0.06 \\
LO 1840 & J05341141+0942079 & 05:34:11.41 & +09:42:07.94 & M6.0 & 0.09 & ... & 1 & 1.24 $\pm$ 0.25 & -2.91 $\pm$ 0.21 \\
LO 2088 & J05341927+0948275 & 05:34:19.27 & +09:48:27.51 & M7.0 & 0.04 & 12.1 $\pm$ 0.5 & 1,5 & -2.06 $\pm$ 1.52 & -1.03 $\pm$ 1.32 \\
LO 7957 & ... & 05:34:36.28 & +09:55:32.20 & M8.0 & ... & ... & 1 & ... & ... \\
LO 2712 & J05343836+0958116 & 05:34:38.36 & +09:58:11.63 & M7 & 0.04 & ... & 2 & 2.85 $\pm$ 2.18 & -5.91 $\pm$ 1.85 \\
LO 2989 & J05344621+0955376 & 05:34:46.21 & +09:55:37.65 & M5.5 & 0.07 & 12.7 $\pm$ 0.4 & 1,5 & -0.63 $\pm$ 1.01 & -1.62 $\pm$ 0.85 \\
LO 2993 & J05344631+1002318 & 05:34:46.32 & +10:02:31.87 & M5.0 & 0.08 & ... & 1 & -1.79 $\pm$ 0.69 & -3.57 $\pm$ 0.60 \\
LO 3360 & J05345639+0955045 & 05:34:56.40 & +09:55:04.46 & M4.0 & 0.19 & 10.7 $\pm$ 0.1 & 1,4 & 0.84 $\pm$ 0.33 & -2.65 $\pm$ 0.25 \\
LO 3506 & J05350015+0952408 & 05:35:00.16 & +09:52:40.88 & M8 & 0.03 & 13.3 $\pm$ 0.7 & 2,5 & 5.26 $\pm$ 3.29 & 0.12 $\pm$ 2.76 \\
LO 3597 & J05350274+0956475 & 05:35:02.74 & +09:56:47.58 & M4.0 & 0.19 & 12.3 $\pm$ 0.2 & 1,4 & 0.57 $\pm$ 0.12 & -2.00 $\pm$ 0.10 \\
LO 3746 & J05350707+0954014 & 05:35:07.07 & +09:54:01.48 & M6 & 0.06 & 12.7 $\pm$ 0.4 & 2,5 & 0.56 $\pm$ 0.62 & -1.84 $\pm$ 0.49 \\
LO 7951 & ... & 05:35:07.95 & +10:00:06.26 & M6.0 & ... & ... & 1 & ... & ... \\
LO 3785 & J05350833+0942537 & 05:35:08.34 & +09:42:53.79 & K4 & 1.04 & 14.5 $\pm$ 0.3 & 2,4 & 1.59 $\pm$ 0.07 & -1.96 $\pm$ 0.06 \\
HD 245185 & J05350960+1001515 & 05:35:09.60 & +10:01:51.43 & A0 & 2.26 & ... & 3 & 0.34 $\pm$ 0.14 & -1.93 $\pm$ 0.10 \\
LO 3887 & J05351112+0957195 & 05:35:11.13 & +09:57:19.58 & M5.0 & 0.09 & 12.8 $\pm$ 0.4 & 1,5 & 0.58 $\pm$ 0.37 & -2.19 $\pm$ 0.33 \\
LO 3942 & J05351255+0953111 & 05:35:12.56 & +09:53:11.14 & M3.0 & 0.30 & 11.2 $\pm$ 0.3 & 1,4 & 1.44 $\pm$ 0.60 & -2.96 $\pm$ 0.46 \\
LO 4021 & J05351533+0948369 & 05:35:15.33 & +09:48:36.96 & M3.0 & 0.33 & 11.6 $\pm$ 0.3 & 1,5 & 0.65 $\pm$ 0.21 & -1.96 $\pm$ 0.16 \\
LO 4111 & J05351792+0956571 & 05:35:17.92 & +09:56:57.20 & M4 & 0.19 & 12.0 $\pm$ 0.3 & 2,5 & 1.13 $\pm$ 0.23 & -1.94 $\pm$ 0.17 \\
LO 4126 & J05351818+0952241 & 05:35:18.18 & +09:52:24.17 & M2.0 & 0.44 & 11.7 $\pm$ 0.4 & 1,5 & 0.50 $\pm$ 0.21 & -1.70 $\pm$ 0.16 \\
LO 4155 & J05351904+0954550 & 05:35:19.05 & +09:54:55.74 & K2 & 1.40 & 12.9 $\pm$ 0.5 & 1,4 & 0.93 $\pm$ 0.08 & -2.19 $\pm$ 0.06 \\
LO 4163 & J05351913+0954424 & 05:35:19.14 & +09:54:42.38 & M4.0 & 0.19 & 11.2 $\pm$ 0.4 & 1,5 & 0.21 $\pm$ 0.20 & -2.03 $\pm$ 0.16 \\
LO 4187 & J05351991+1002364 & 05:35:19.92 & +10:02:36.51 & M1.0 & 0.50 & 9.2 $\pm$ 0.3 & 1,6 & 2.66 $\pm$ 0.09 & -2.76 $\pm$ 0.07 \\
LO 4255 & J05352151+0953291 & 05:35:21.52 & +09:53:29.21 & M5 & 0.10 & 12.1 $\pm$ 0.3 & 2,5 & 0.12 $\pm$ 0.36 & -1.47 $\pm$ 0.35 \\
LO 4363 & J05352440+0953519 & 05:35:24.41 & +09:53:51.94 & M5.5 & 0.07 & 12.3 $\pm$ 0.4 & 1,5 & 0.35 $\pm$ 0.64 & -0.18 $\pm$ 0.50 \\
LO 4407 & J05352536+1008383 & 05:35:25.36 & +10:08:38.25 & M3 & 0.29 & 11.2 $\pm$ 0.2 & 2,4 & 1.71 $\pm$ 0.54 & -4.05 $\pm$ 0.43 \\
LO 4520 & J05352846+1002275 & 05:35:28.46 & +10:02:27.44 & M3.5 & 0.25 & ... & 1 & 0.62 $\pm$ 0.16 & -1.56 $\pm$ 0.12 \\
LO 4531 & J05352877+0954101 & 05:35:28.78 & +09:54:10.08 & M5.5 & 0.07 & 11.9 $\pm$ 0.3 & 1,5 & 1.51 $\pm$ 0.71 & -2.74 $\pm$ 0.62 \\
LO 4817 & J05353722+0956517 & 05:35:37.23 & +09:56:51.72 & M4 & 0.20 & ... & 2 & 0.82 $\pm$ 0.20 & -2.28 $\pm$ 0.16 \\
LO 4916 & J05353984+0953240 & 05:35:39.85 & +09:53:24.06 & M6.5 & 0.05 & 11.8 $\pm$ 0.7 & 1,5 & ... & ... \\
LO 5267 & J05355094+0938567 & 05:35:50.95 & +09:38:56.69 & M4 & 0.19 & ... & 1 & 0.87 $\pm$ 0.28 & -1.77 $\pm$ 0.21 \\
LO 5447 & J05355585+0956217 & 05:35:55.86 & +09:56:21.75 & M1.5 & 0.24 & 10.8 $\pm$ 0.2 & 1,4 & 0.98 $\pm$ 0.13 & -1.87 $\pm$ 0.10 \\
LO 5679 & J05360288+0942074 & 05:36:02.88 & +09:42:07.50 & M1 & 0.47 & 11.4 $\pm$ 0.1 & 2,4 & 2.16 $\pm$ 0.09 & -2.28 $\pm$ 0.06 \\
LO 5916 & J05360981+0942370 & 05:36:09.81 & +09:42:37.02 & M6.0 & 0.05 & 11.8 $\pm$ 0.5 & 1,5 & 0.58 $\pm$ 1.07 & -2.09 $\pm$ 0.98 \\
LO 6191 & J05361810+0952254 & 05:36:18.11 & +09:52:25.41 & M6.0 & 0.06 & 11.9 $\pm$ 0.5 & 1,5 & 1.39 $\pm$ 0.82 & -3.26 $\pm$ 0.61 \\
LO 6866 & J05363804+0940509 & 05:36:38.07 & +09:40:50.18 & K7 & 0.67 & 11.9 $\pm$ 0.2 & 2,4 & 0.60 $\pm$ 0.12 & -1.79 $\pm$ 0.11 \\
LO 6886 & J05363861+0935052 & 05:36:38.61 & +09:35:05.22 & M3 & 0.29 & 11.5 $\pm$ 0.2 & 2,4 & 0.83 $\pm$ 0.44 & -2.14 $\pm$ 0.36 \\
LO 7402 & J05365309+0941556 & 05:36:53.09 & +09:41:55.67 & K4 & 1.05 & 13.6 $\pm$ 0.1 & 2,4 & 1.96 $\pm$ 0.06 & -1.97 $\pm$ 0.05 \\
LO 7490 & J05365533+0946479 & 05:36:55.34 & +09:46:47.94 & M3 & 0.33 & ... & 1 & 1.15 $\pm$ 0.18 & -2.54 $\pm$ 0.17 \\
LO 7528 & J05365617+0931227 & 05:36:56.17 & +09:31:22.70 & M1.5 & 0.42 & 9.5 $\pm$ 0.2 & 1,4 & ... & ...
\enddata 
\tablenotetext{}{$^\dagger$References for stellar spectral types (SpT) and radial velocities ($v_{\rm LSR}$): (1) \cite{Bayo2011}, (2) \cite{Hernandez2010}, (3) \cite{Hernandez2009}, (4) \cite{Kounkel2018}, (5) \cite{Maxted2008}, (6) \cite{Sacco2008}. }\end{deluxetable*}


\section{Sample} \label{sec:sample}

\subsection{Disk Sample Selection} \label{sec:disks}

Like most SFRs, the disk census in $\lambda$~Orionis is based on targeted {\it Spitzer} observations, which can identify stars exhibiting excess emission above the stellar photosphere at near-infrared \citep[IRAC 3.6, 4.5, 5.8, and 8.0~$\mu$m;][]{Fazio2004} and/or mid-infrared \citep[MIPS 24~$\mu$m;][]{Rieke2004} wavelengths where dust emits efficiently. \cite{Hernandez2009} used {\it Spitzer} data to study the intermediate-mass population in $\lambda$~Orionis, finding 29 members earlier than F5 but only 10 bearing disks; they classified the nine sources with moderate infrared excess as debris disks and the one source with large infrared excess as an optically thick disk. \cite{Hernandez2010} then studied the lower-mass population in $\lambda$~Orionis, finding 436 members down to the substellar limit but only 49 with disks; they grouped the disks according to their spectral energy distributions (SEDs)---optically thick disks had the largest excesses, evolved disks had smaller excesses, and (pre-)transition disks exhibited signs of inner disk clearings. We exclude from our sample the nine debris disks, as these are likely second-generation dust disks \citep{Wyatt2008}, resulting in an initial sample of 50 primordial (or protoplanetary) disks. We use the same naming conventions as \cite{Hernandez2009,Hernandez2010} in this work.

As shown in Figure~\ref{fig:pm}, the region surveyed by {\it Spitzer} (white box) only covers $\sim$3~pc from the cluster center. Recent kinematic studies of the Orion Complex using astrometric data from {\it Gaia} \citep{Gaia2016} Data Release 2 \cite[DR2; ][]{DR22018} combined with spectroscopic data from the Apache Point Observatory Galactic Evolution Experiment \cite[APOGEE;][]{Majewski2017} revealed a population of radially expanding $\lambda$ Orionis members that extend well beyond this area \citep{Kounkel2018}. Surveys for disks in these outer regions have not yet been conducted, making it possible that our ALMA sample based solely on {\it Spitzer} data is incomplete. Nevertheless, the low protoplanetary disk fraction in $\lambda$~Orionis inferred from the {\it Spitzer} data is consistent with the $\sim$5~Myr age of the cluster, as it follows the well-known exponential decline in disk frequency with age \citep[e.g., see Figure~14 in][]{Hernandez2007}. Moreover, the issue of a potentially incomplete disk census when using targeted {\it Spitzer} observations for the sample selection is a general (albeit moderate) problem facing the ALMA disk demographic literature, as {\it Gaia} continues to reveal missed or interloping stellar populations in young SFRs \cite[e.g.,][]{Manara2018, Galli2020,LE2020}. We also note that the radially expanding population missed by {\it Spitzer} is unlikely to represent a younger population formed as a consequence of the supernova and/or feedback from the massive OB stars, as \cite{DM1999,DM2001} found no evidence for triggered or sequential star formation in the region. 

We identify interlopers in the {\it Spitzer} sample by using distances from {\it Gaia} DR2 to find contaminant background sources---a method that has been previously applied to refine membership in the Lupus clouds \citep{Manara2018}. While the individual {\it Gaia} DR2 parallaxes of $\lambda$ Orionis members remain imprecise due to the cluster's distance, they are sufficient for identifying clear interlopers. We therefore remove five sources (LO~1310, 2357, 2404, 5042, and 7517) from our sample, as the lower bounds of their estimated {\it Gaia} DR2 distances from \cite{BJ2018} are $>480$~pc, making them likely background sources. Indeed, LO~1310 and LO~2357 have radial velocities in the local standard of rest (LSR) reference frame of $v_{\rm LSR} = 1.3$ and $-35$~km~s$^{-1}$, respectively, which differ significantly from the average cluster value of 12~km~s$^{-1}$ \citep{Kounkel2018}; LO~2404, 5042, and 7517 do not have known radial velocities. We also remove LO~3710, found to be a non-member by \cite{Bayo2011} due to a discrepant surface gravity. Of these interlopers, only LO~7517 is detected by our ALMA survey at marginal (3.4$\sigma$) significance; given its estimated {\it Gaia} DR2 distance of 3744$_{-336}^{+405}$~pc, it is likely a background galaxy. 

Table~\ref{tab-main} gives our final sample of the 44 protoplanetary disks analyzed in the remainder of this work.

\subsection{Host Star Properties} \label{sec:stars}

The available host star properties for our disk sample are provided in Table~\ref{tab-main}. Stellar spectral types (SpT) were mostly determined by \cite{Bayo2011} from moderate-resolution optical and near-infrared spectra. However, 14 stars have only photometric spectral types from \cite{Hernandez2010}, who interpolated $R-J$ colors onto the spectral type sequence using the standard $R-J$ colors from \cite{KH1995}. Although the $R-J$ colors were not corrected for reddening, the reddening toward $\lambda$~Orionis is low at $E(B-V)\approx0.12$ \citep{DS1994} and these photometric spectral types match well (typically $\pm$1 spectral sub-type) to the spectroscopically determined values from \cite{Bayo2011} when the samples overlap. 

The host star proper motions in right ascension ($\mu_{RA}$) and declination ($\mu_{Dec}$) are from {\it Gaia} DR2. Radial velocities are also provided and taken from various literature sources, translated into the LSR reference frame ($v_{\rm LSR}$) using the source coordinates. The source coordinates in right ascension (R.A.) and declination (Decl.) are the fitted positions for our ALMA detections (Section~\ref{sec:cont}) or the Two Micron All-Sky Survey \citep[2MASS;][]{TWOMASS2006AJ} positions for the non-detections.

As previously mentioned, the $\lambda$ Orionis cluster is too distant for reliable individual {\it Gaia} DR2 parallaxes. This is further complicated for young, disk-hosting stars whose variability from disk scattered light and/or surrounding nebulosity can result in poor fits to the {\it Gaia} DR2 single-star astrometric model, leading to higher astrometric noise. Thus we do not provide distance estimates for each source in our sample, but rather rely on the average distance of the non-disk-bearing members in $\lambda$~Orionis from {\it Gaia} DR2, which is well-constrained to 404$\pm$4~pc \citep{Kounkel2018} and consistent with earlier estimates of 450$\pm$50~pc based on main-sequence fitting to the massive OB stars \citep{DM2001,Mathieu2015}. We therefore adopt a distance of 400~pc for $\lambda$~Orionis in the remainder of this work.

We estimate stellar masses ($M_\star$) from these SpT values and 2MASS $J$-band magnitudes, following the methods of \cite{Ansdell2017}. Each target is placed on the Hertzsprung–Russel (HR) diagram by converting SpT to stellar effective temperature and $J$-band magnitude to stellar luminosity using the relations from \cite{HH2015} and a distance of 400~pc. Estimates of $M_\star$ are then found by comparing the positions on the HR diagram to the evolutionary models of \cite{Baraffe2015}. For the one intermediate-mass star, HD~245185, we instead use the evolutionary models of \cite{Siess2000}. We do not provide $M_\star$ estimates for the two sources without 2MASS data (LO~7957 and LO~7951). The typical $M_\star$ uncertainties, propagated from the uncertainties on SpT and $J$-band magnitude, are 0.1--0.2~$M_\odot$.

\begin{figure*}[!ht]
\begin{center}
\includegraphics[width=18.2cm]{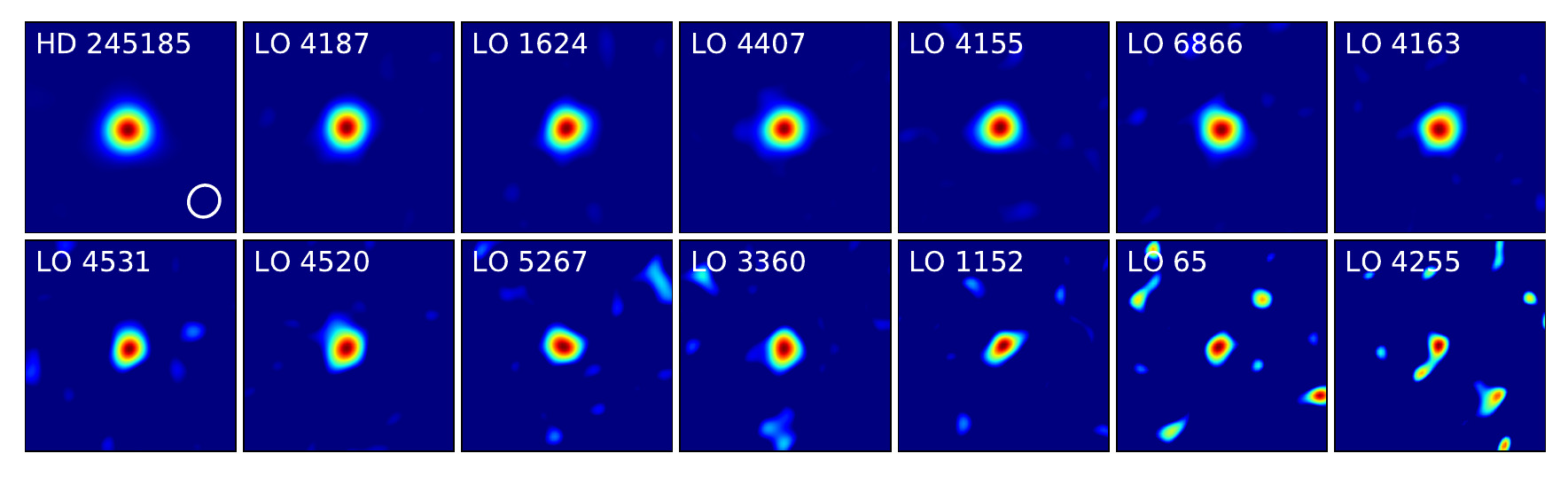}
\caption{ALMA 1.25~mm continuum images of the 14 detected disks in our $\lambda$ Orionis sample, ordered by decreasing flux density (as reported in Table~\ref{tab-alma}). The 2\as{}$\times$2\as{} images are centered on the source, scaled to their maximum value, and clipped below 1.5$\sigma$ for clarity. The typical beam size of $\sim$0.3\as{} is shown in the first panel by the white ellipse.}
\label{fig:stamps}
\end{center}
\end{figure*}


\section{ALMA Observations \& Results} 
\label{sec:alma}

\subsection{ALMA Observations}
\label{sec:obs}

The ALMA observations used in this work were taken in Cycle 5 under program 2017.1.00466.S (PI: Ansdell). The program allowed for a range of array configurations to maximize the probability of survey completion, and the observations were conducted over 13 execution blocks (see  Appendix~\ref{app:obslog} for information on the array configuration and weather conditions for each execution block). All sources in our sample were observed during each execution block, thus the sensitivities and synthesized beams are uniform across the sample.

The spectral setup was identical for each execution block: four spectral windows were centered on 247.96, 245.46, 232.97, and 230.51~GHz, each with usable bandwidths of 1.88~GHz, for a bandwidth-weighted mean continuum frequency of 239.36~GHz (1.25~mm). The last spectral window covered the $^{12}$CO~$J=2-1$ transition at moderate ($\sim$1~km~s$^{-1}$) spectral resolution to preserve the maximum possible continuum bandwidth while allowing for the possibility of detecting molecular gas emission. 

On-source integration times were $\sim$6~min per source, chosen so the $3\sigma$ dust mass constraints would be comparable to the limits reached in the ALMA surveys of more nearby SFRs \citep[e.g.,][]{Barenfeld2016,Ansdell2016,Pascucci2016}, assuming a linear relation between millimeter flux and dust mass \citep[e.g.,][]{Hildebrand1983}. Data were pipeline calibrated by NRAO staff using the Common Astronomy Software Applications ({\tt CASA}) package \citep{CASA2007} version~5.4.0. The pipeline included flux, bandpass, and gain calibrations (see Appendix~\ref{app:obslog} for a list of the calibrators). We assume an absolute flux calibration uncertainty of 10\%, similar to other ALMA disk surveys \cite[e.g.,][]{Barenfeld2016,Ansdell2016, Cazzoletti2019}.

\subsection{ALMA Continuum Results}
\label{sec:cont}

We create continuum images from the calibrated visibilities by averaging over the continuum channels using the {\tt split} task in {\tt CASA}, then cleaning with a Briggs robust weighting parameter of $+0.5$ using the {\tt tclean} task. This results in a median continuum rms of 34~$\mu$Jy and beam size of $0.31$\as$\times$0.29\as{}. We do not perform self-calibration as the detected sources are faint, with a median signal-to-noise ratio of 15 (see Table~\ref{tab-alma}). 

In most cases, we measure continuum flux densities by fitting point source models to the visibility data with the {\tt uvmodelfit} task in {\tt CASA}. The point source model has three free parameters: integrated flux density ($F_{\rm 1.25mm}$), right ascension offset from the phase center ($\Delta\alpha$), and declination offset from the phase center ($\Delta\delta$). For the non-detections, we fix $\Delta\alpha$ and $\Delta\delta$ to zero when running {\tt uvmodelfit} to avoid spurious detections (the phase offsets are typically only 0.05\as{} for the detections, much smaller than the beam size). In three cases, the sources are resolved, therefore we use an elliptical Gaussian model instead, which has three additional parameters: full-width-half-max along the major axis ($a$), aspect ratio of the axes ($r$), and position angle (${\rm P.A.}$). With the underlying assumption that
these models describe the data appropriately, we multiply the uncertainties on all the fitted parameters by the factor needed to produce a reduced $\chi^{2}$ of 1 (typically a factor of two).

Table~\ref{tab-alma} reports the $F_{\rm 1.25mm}$ values for all sources, along with their statistical uncertainties (i.e., not including the 10\% flux calibration error), while Figure~\ref{fig:stamps} shows the continuum images for the $\ge3\sigma$ detections. Only 14 of the 44 sources are detected at $\ge3\sigma$ significance and only 3 of the detections are marginally resolved. We conservatively identify resolved sources as those where the ratio of $a$ to its uncertainty is greater than five. The resolved sources are HD~245185, LO~4187, and LO~4407 with the following fitted elliptical Gaussian parameters, respectively: $a = 0.203$\as{}$\pm$0.002\as{}, 0.113\as{}$\pm$0.015\as{}, and 0.123\as{}$\pm$0.022\as{}; $r = 0.807$$\pm$0.006, 0.710$\pm$0.196, and 0.508$\pm$0.380; ${\rm P.A.} = 70.1$\degrees{}$\pm$0.9\degrees{}, $-1.3$\degrees{}$\pm$19.5\degrees{}, and 85.5\degrees{}$\pm$18.7\degrees{}.

We stack the images of the 30 non-detections to constrain the average continuum flux for the individually undetected sources, finding a tentative detection of $0.019\pm0.006$~mJy (3.2$\sigma$) in the stacked image. We verify this result by calculating the mean continuum flux density and standard error on the mean for the 30 non-detections using the values in Table~\ref{tab-alma}, which gives $0.020\pm0.005$~mJy (4.0$\sigma$).

\begin{deluxetable}{lrr} 
\tabletypesize{\footnotesize} 
\centering 
\tablewidth{0pt} 
\tablecaption{ALMA Disk Fluxes \label{tab-alma}} 
\tablecolumns{3}  
\tablehead{ 
 \colhead{Source} 
&\colhead{$F_{\rm 1.25mm}$} 
&\colhead{$F_{\rm 12CO}$} 
\\ 
 \colhead{} 
&\colhead{(mJy)} 
&\colhead{(mJy km s$^{-1}$)} 
} 
\renewcommand{\arraystretch}{1.18} 
\startdata 
LO 65 & 0.117 $\pm$ 0.030 & $<$ 25 \\
LO 1079 & 0.033 $\pm$ 0.030 & $<$ 24 \\
LO 1152 & 0.159 $\pm$ 0.028 & $<$ 24 \\
LO 1359 & 0.053 $\pm$ 0.028 & $<$ 23 \\
LO 1589 & 0.053 $\pm$ 0.028 & $<$ 24 \\
LO 1624 & 1.179 $\pm$ 0.028 & $<$ 24 \\
LO 1840 & -0.008 $\pm$ 0.030 & $<$ 24 \\
LO 2088 & -0.005 $\pm$ 0.028 & $<$ 24 \\
LO 7957 & 0.075 $\pm$ 0.028 & $<$ 24 \\
LO 2712 & -0.033 $\pm$ 0.028 & $<$ 25 \\
LO 2989 & 0.076 $\pm$ 0.028 & $<$ 23 \\
LO 2993 & 0.008 $\pm$ 0.030 & $<$ 24 \\
LO 3360 & 0.234 $\pm$ 0.028 & $<$ 24 \\
LO 3506 & 0.016 $\pm$ 0.030 & $<$ 25 \\
LO 3597 & -0.063 $\pm$ 0.030 & $<$ 25 \\
LO 3746 & 0.047 $\pm$ 0.030 & $<$ 24 \\
LO 7951 & 0.038 $\pm$ 0.028 & $<$ 25 \\
LO 3785 & 0.035 $\pm$ 0.034 & $<$ 28 \\
HD 245185 & 33.955 $\pm$ 0.056 & 1915 $\pm$ 51 \\
LO 3887 & 0.024 $\pm$ 0.028 & $<$ 23 \\
LO 3942 & -0.001 $\pm$ 0.028 & $<$ 24 \\
LO 4021 & 0.032 $\pm$ 0.028 & $<$ 24 \\
LO 4111 & -0.012 $\pm$ 0.028 & $<$ 24 \\
LO 4126 & 0.004 $\pm$ 0.028 & $<$ 23 \\
LO 4155 & 0.927 $\pm$ 0.030 & 27 $\pm$ 8 \\
LO 4163 & 0.481 $\pm$ 0.028 & $<$ 24 \\
LO 4187 & 1.846 $\pm$ 0.037 & 113 $\pm$ 21 \\
LO 4255 & 0.099 $\pm$ 0.028 & $<$ 24 \\
LO 4363 & 0.032 $\pm$ 0.028 & $<$ 23 \\
LO 4407 & 1.067 $\pm$ 0.037 & 29 $\pm$ 8 \\
LO 4520 & 0.340 $\pm$ 0.028 & $<$ 25 \\
LO 4531 & 0.375 $\pm$ 0.028 & $<$ 22 \\
LO 4817 & 0.023 $\pm$ 0.028 & $<$ 24 \\
LO 4916 & 0.017 $\pm$ 0.028 & $<$ 24 \\
LO 5267 & 0.257 $\pm$ 0.028 & $<$ 24 \\
LO 5447 & -0.002 $\pm$ 0.028 & $<$ 23 \\
LO 5679 & 0.054 $\pm$ 0.028 & $<$ 24 \\
LO 5916 & 0.011 $\pm$ 0.028 & $<$ 23 \\
LO 6191 & 0.009 $\pm$ 0.028 & $<$ 24 \\
LO 6866 & 0.623 $\pm$ 0.028 & 80 $\pm$ 14 \\
LO 6886 & 0.006 $\pm$ 0.028 & $<$ 23 \\
LO 7402 & 0.017 $\pm$ 0.030 & $<$ 23 \\
LO 7490 & 0.030 $\pm$ 0.028 & $<$ 25 \\
LO 7528 & 0.031 $\pm$ 0.028 & $<$ 25
\enddata 
\end{deluxetable}

\subsection{ALMA CO Results}
\label{sec:co}

We extract $^{12}$CO channel maps from the calibrated visibilities by first subtracting the continuum using the {\tt uvcontsub} task in {\tt CASA}. To search for $^{12}$CO emission, we follow the general procedure of \cite{Ansdell2017}. In short, we extract an initial spectrum for each source to identify candidate detections with emission exceeding 3$\times$ the channel rms near the expected $v_{\rm LSR}$ of the cluster \cite[12~km~s$^{-1}$;][]{Kounkel2018}. These candidates are visually inspected and any emission is cleaned with a Briggs robust weighting parameter of $+0.5$ using {\tt tclean}. Zero-moment maps are then created by summing the channels $\pm3$~km~s$^{-1}$ from the systemic velocity unless clear emission was seen beyond these limits. The integrated $^{12}$CO line fluxes ($F_{\rm 12CO}$) are measured using a curve-of-growth aperture photometry method, with errors ($E_{\rm 12CO}$) estimated by taking the standard deviation of fluxes measured within the same-sized aperture placed randomly within the field of view but away from the source. 

{Only five sources (HD~245185, LO~4155, LO~4187, LO~4407, LO~6866) are detected with $F_{\rm 12CO} \ge 3 \times E_{\rm 12CO}$, though two of these (LO~4155, LO~4407) are marginal (3--4$\sigma$) detections. One other source, LO~1624, shows emission at $\sim$5$\times$ the rms in the single channel at its known $v_{\rm LSR}$, but is undetected in its zero-moment map as summing over several channels dilutes the emission. For the non-detections, we construct zero-moment maps by integrating $\pm$3~km~s$^{-1}$ from their known $v_{\rm LSR}$ when available (Table~\ref{tab-main}), else from the average $v_{\rm LSR}$ of the cluster \cite[12~km~s$^{-1}$;][]{Kounkel2018}. Table~\ref{tab-alma} gives the $F_{\rm 12CO}$ values for the detections and upper limits of 3$\times$ the rms in the zero-moment maps for the non-detections. Figure~\ref{fig:gasdet} shows the zero-moment and first-moment maps for the five detections, while Appendix~\ref{app:co} presents the $^{12}$CO spectra for all sources in our sample.

Figure~\ref{fig:gasdist} then shows ALMA Band~6 $^{12}$CO emission as a function of the Band~6 continuum for the continuum-detected disks in the evolved $\lambda$~Orionis cluster (this work), the young Lupus clouds \citep{Ansdell2018}, and the middle-aged $\sigma$~Orionis cluster (\citealt{Ansdell2017}; Ansdell et al., in prep), all scaled to 150~pc. We do not show other notable ALMA disk surveys, as they were conducted in Band~7 \cite[e.g.,][]{Barenfeld2016, Pascucci2016} or do not have published Band~6 $^{12}$CO fluxes \cite[e.g.,][]{Cieza2019}. Interestingly, the roughly linear correlation between continuum and $^{12}$CO flux holds over the first $\sim$5~Myr of disk evolution, implying that the lack of gas detections in our $\lambda$~Orionis survey can be explained by the low continuum emission.

\begin{figure}
\begin{center}
\includegraphics[width=8.2cm]{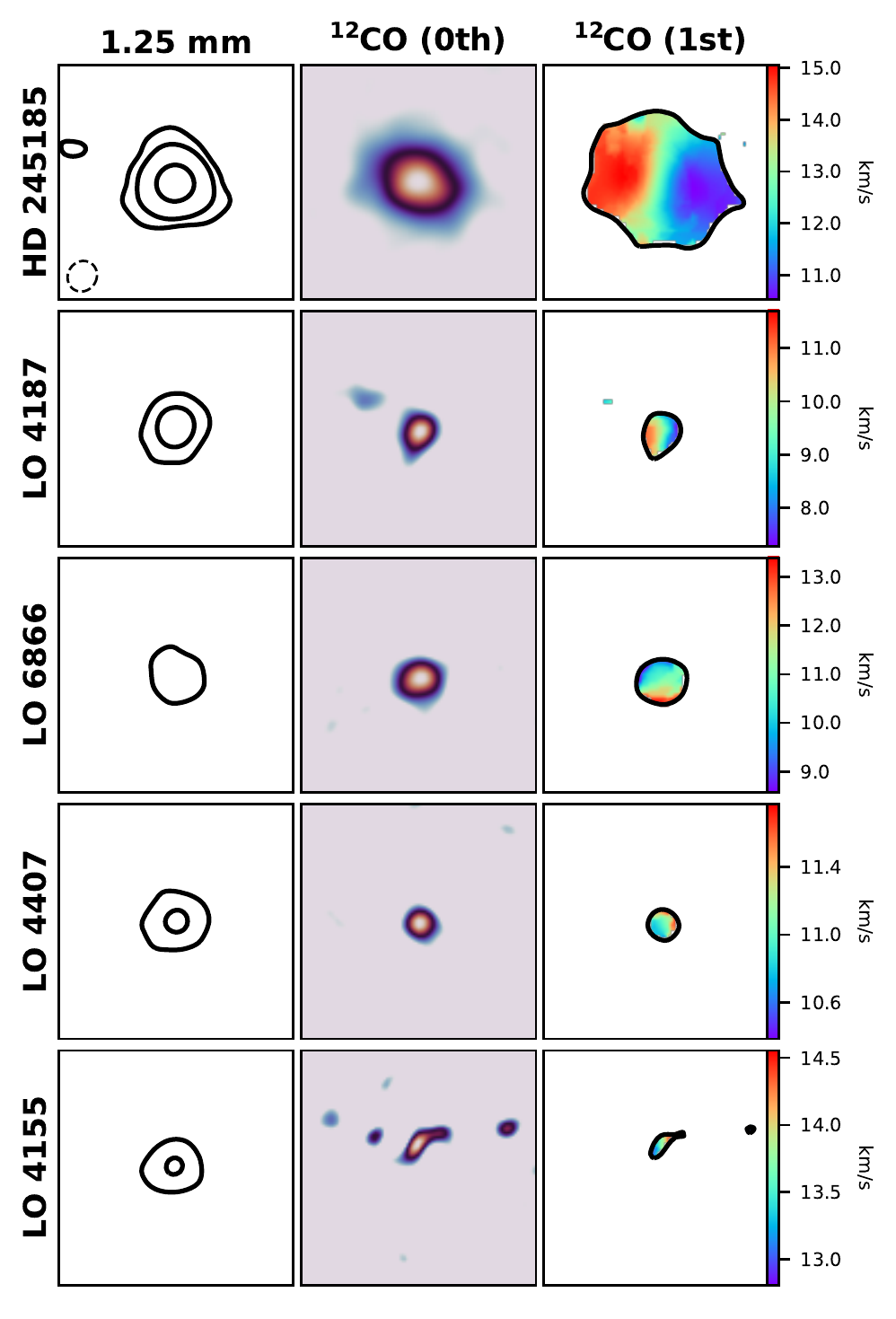}
\caption{Our ALMA Band~6 observations of the five $^{12}$CO detections in our sample (Section~\ref{sec:co}). The first column shows the 1.25~mm continuum maps with 4$\sigma$, 20$\sigma$, and 150$\sigma$ contours. The middle column shows the $^{12}$CO zero-moment maps, scaled to their maximum value and clipped below 2$\sigma$ for clarity. The final column shows the $^{12}$CO first-moment maps within the 3$\sigma$ contours of the zero-moment maps. Images are 2\as{}$\times$2\as{} and the typical beam size is given in the first panel by the dashed ellipse.}
\label{fig:gasdet}
\end{center}
\end{figure}

\begin{figure}
\begin{center}
\includegraphics[width=8.5cm]{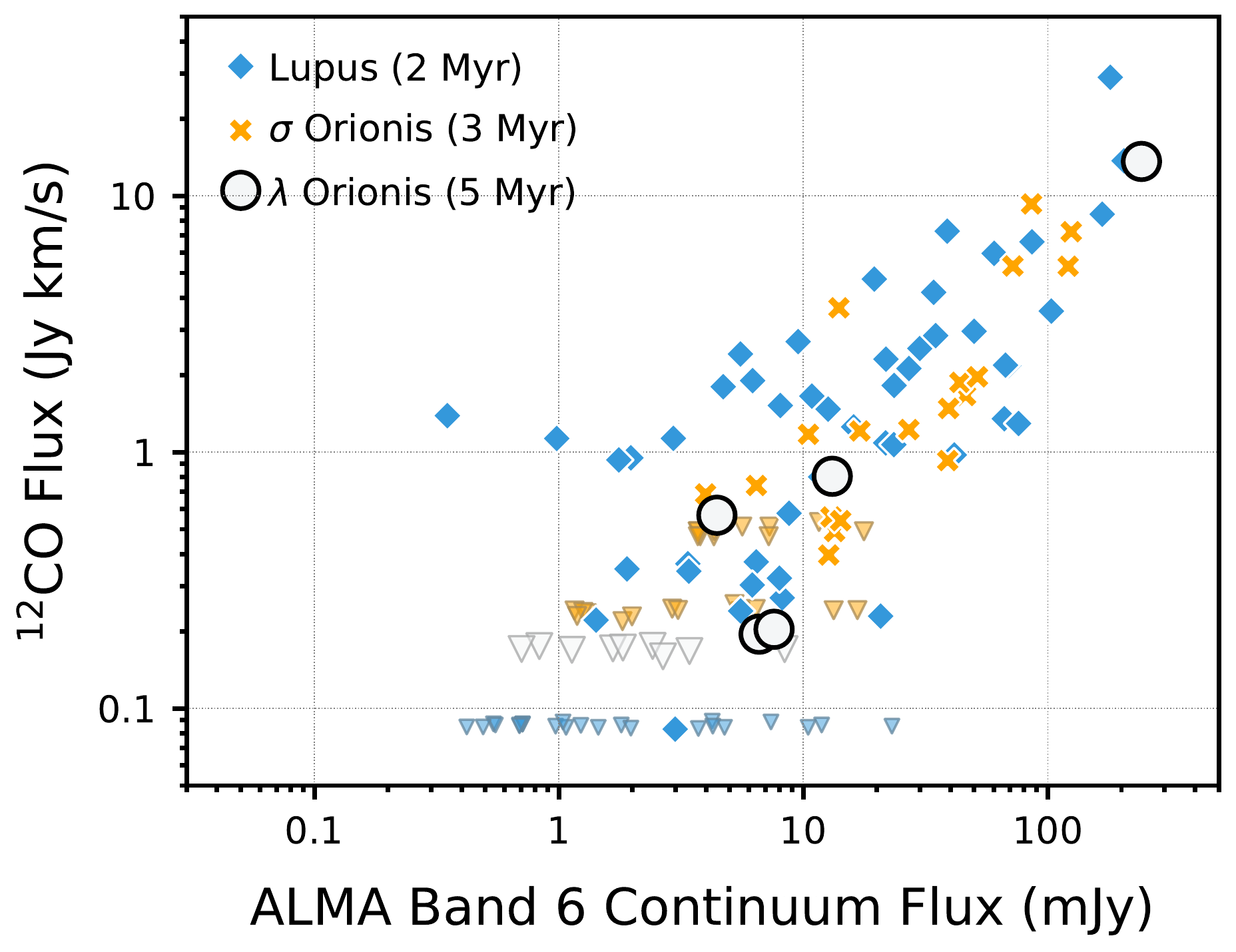}
\caption{$^{12}$CO flux as a function of continuum flux in ALMA Band~6 for the continuum-detected disks in Lupus \citep{Ansdell2018}, $\sigma$~Orionis \citep[][Ansdell et al., in prep]{Ansdell2017}, and $\lambda$~Orionis (this work). Fluxes are normalized to 150~pc and downward-facing triangles are 3$\sigma$ upper limits on $^{12}$CO flux. Approximate ages of each region are provided for reference. The four $\lambda$~Orionis $^{12}$CO detections follow the roughly linear correlation seen in younger regions, and the lingering bright disk in $\lambda$~Orionis is HD~245185.}
\label{fig:gasdist}
\end{center}
\end{figure}

This roughly linear correlation between the (somewhat) optically thin millimeter continuum flux and the optically thick $^{12}$CO flux is potentially due to more massive dust disks having more extended gas disks \citep[e.g.,][]{Barenfeld2016}. Moreover, since there is an observed linear relationship between the millimeter continuum luminosity ($L_{\rm mm}$) and emitting surface area ($R_{\rm mm}^2$) of protoplanetary disks \citep{Tripathi2017,Andrews2018}, and Figure~\ref{fig:gasdist} implies that the $^{12}$CO luminosity ($L_{\rm CO}$) is proportional to $L_{\rm mm}$, while we also expect $L_{\rm CO}$ to be proportional to the $^{12}$CO emitting surface area ($R_{\rm CO}^2$) if it is optically thick, we can predict that $R_{\rm CO} \propto R_{\rm mm}$, which is indeed seen in observations \citep{Ansdell2018,Trapman2020}. 

\newpage


\section{Disk Properties} 
\label{sec:properties}

\subsection{Dust Masses} 
\label{sec:mass}

Under the simplified assumption that dust emission from a protoplanetary disk at millimeter wavelengths is optically thin and isothermal, the observed continuum flux density at a given frequency ($F_\nu$) can be directly related to the mass of the emitting dust ($M_{\rm dust}$), as established in \cite{Hildebrand1983}:

\begin{equation}
M_{\rm dust} =\frac{F_{\nu}d^{2}}{\kappa_{\nu}B_{\nu}(T_{\rm dust})} \approx 4.1 \times F_{1.25 {\rm mm}},
\label{eqn-mass}
\end{equation}

\begin{figure*}
\begin{center}
\includegraphics[width=18cm]{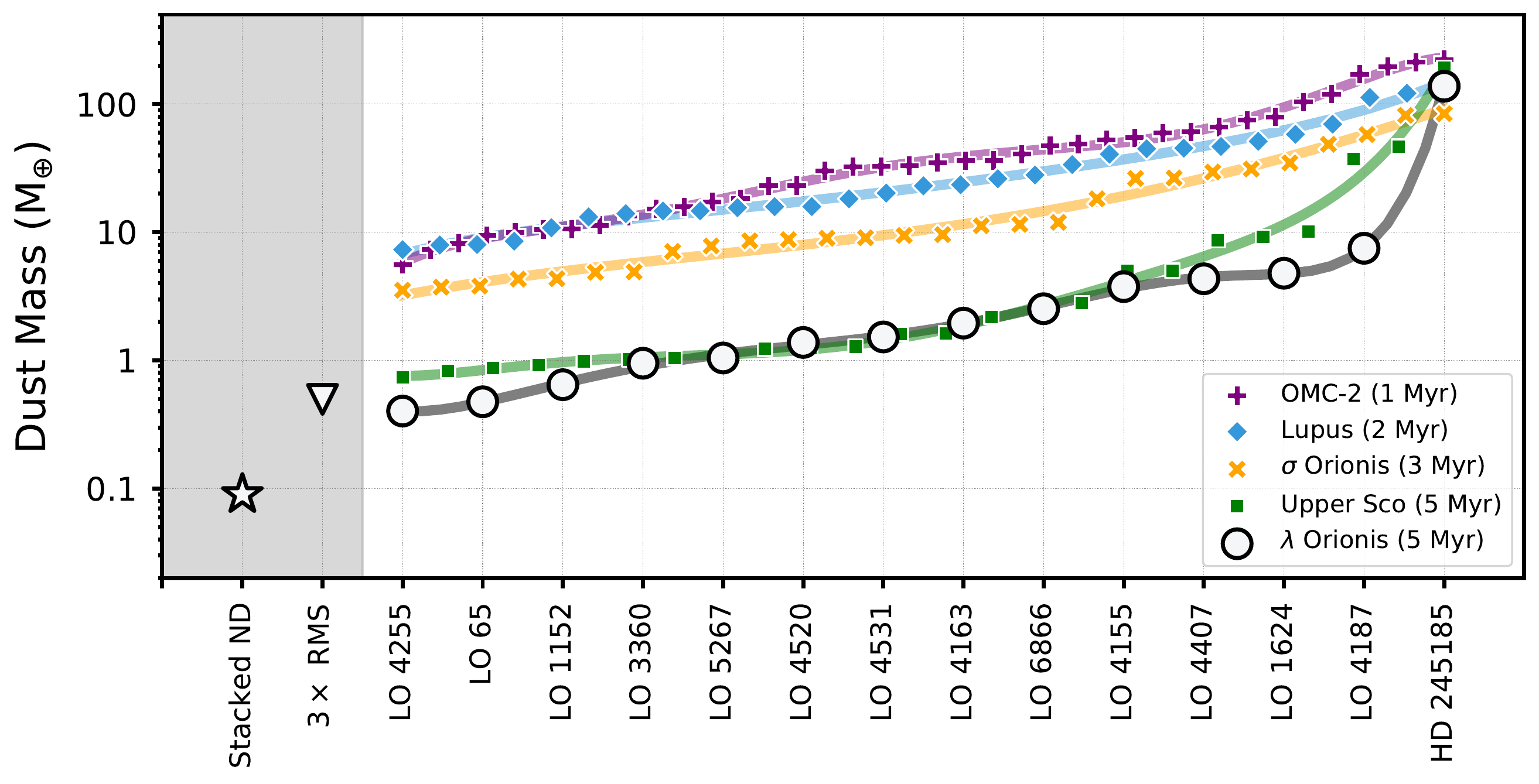}
\caption{Dust masses of the 14 continuum-detected disks in our $\lambda$~Orionis survey (white circles) in increasing order (source names are given on the $x$ axis). The white downward triangle is the median 3$\sigma$ upper limit for individual non-detections (ND), while the white star shows their detected average dust mass from our stacking analysis (Section~\ref{sec:mass}). Comparable disk dust mass populations (see Section~\ref{sec:compare}) from OMC-2 (purple crosses), Lupus (blue diamonds), $\sigma$~Orionis (orange x's), and Upper Sco (green squares) are also shown, illustrating the smooth distribution in the younger regions in contrast with the steep rise toward the high-mass outlier disks in the older populations (see Sections~\ref{sec:compare} and \ref{sec:outliers}). The colored solid lines are polynomial fits to guide the eye for each region; approximate ages of each region are also provided for reference.}
\label{fig:mdust}
\end{center}
\end{figure*}

where $M_{\rm dust}$ is in Earth masses and $F_{\rm 1.25 mm}$ is in mJy. Here we use $B_{\nu}(T_{\rm dust})$ as the Planck function for a characteristic dust temperature of $T_{\rm dust}=20$~K, the median for Taurus disks \citep{Andrews2005}. We take the dust grain opacity, $\kappa_{\nu}$, as 2.3~cm$^{2}$ g$^{-1}$ at 230~GHz and use an opacity power-law index of $\beta_{\rm d} =1.0$ \citep{Beckwith1990}; these are the same assumptions as in \cite{Ansdell2016,Ansdell2017,Ansdell2018} and \cite{Cieza2019} but differ slightly from some previous ALMA disk surveys \citep{Pascucci2016,Barenfeld2016}, which use $\beta=0.4$. For the distance, $d$, we use the cluster's average {\it Gaia} DR2 value of 400~pc (Section~\ref{sec:stars}), and take the $F_{\rm 1.25mm}$ measurements from Table~\ref{tab-alma}. 

With this approach, the median $M_{\rm dust}$ of the continuum detections in our $\lambda$~Orionis survey is only $\sim$2~$M_\oplus$. These dust masses may be underestimated, however, if the observed millimeter continuum emission is (partially) optically thick \citep[e.g.,][]{Andrews2005, Zhu2019} and/or if the temperature of the dust in the outer disk (where we assume most of the disk mass is located) is lower than 20~K. Nevertheless, employing this simplified relation with these caveats in mind provides the most practical approach given the faint and unresolved emission from the disks in our observations.

Figure~\ref{fig:mdust} shows the $M_{\rm dust}$ estimates for the continuum detections in our survey in increasing order. It also includes the median 3$\sigma$ upper limit of $\sim$0.4~$M_\oplus$ for the individual non-detections in our survey, as well as the mean detection of $\sim$0.08~$M_\oplus$ found when stacking these non-detections (Section~\ref{sec:cont}).

\subsection{Comparisons to Other Regions} 
\label{sec:compare}

At $\sim$5~Myr old, $\lambda$~Orionis provides an important point of comparison to the several younger disk populations, as well as the similarly aged Upper Sco disk population, that have been previously surveyed by ALMA. Figure~\ref{fig:kme} compares the $M_{\rm dust}$ cumulative distribution for $\lambda$~Orionis (this work) to that of OMC-2 \citep{Terwisga2019}, Lupus \citep{Ansdell2018}, $\sigma$~Orionis \citep{Ansdell2017}, and Upper Sco \citep{Barenfeld2016}. The $M_{\rm dust}$ values are uniformly calculated using Equation~\ref{eqn-mass} with the reported wavelengths of the surveys and typical {\it Gaia} DR2 distances of $\sim$150~pc for Lupus and Upper Sco and $\sim$400~pc for $\sigma$~Orionis and OMC-2. For the approximate ages of the regions, we adopt the values used in the ALMA surveys or reported in more recent analyses of the protoplanetary disk populations \citep[e.g.,][]{Andrews2020}. The cumulative distributions are constructed using the Kaplan-Meier Estimator (with the Python {\tt lifelines} package; \citealt{lifelines}) to account for upper limits, as in previous works \cite[e.g.,][]{Barenfeld2016, Ansdell2017,Cieza2019,vanTerwisga2019,Cazzoletti2019}. 

Figure~\ref{fig:kme} illustrates that $\lambda$~Orionis follows the general decay in the overall disk dust mass population with age that has been previously reported \citep[e.g.,][]{Ansdell2016,Barenfeld2016,Pascucci2016,Cieza2019,vanTerwisga2019}. Moreover, the $M_{\rm dust}$ distribution in $\lambda$~Orionis is statistically indistinguishable from that of the similarly aged Upper Sco association. Both of these findings suggest that, if a supernova did occur relatively recently in $\lambda$~Orionis, it did not have a significant impact on disk dust mass evolution in the region. We discuss this further in Section~\ref{sec:sn}.

A reliable comparison of the $M_{\rm dust}$ distributions in Figure~\ref{fig:kme} requires confirming that the regions have similar $M_{\star}$ populations, due to the known correlation between disk dust mass and stellar mass \cite[e.g.,][]{Andrews2013, Ansdell2016, Barenfeld2016, Pascucci2016}. Two-sample tests have previously demonstrated that the stellar mass populations in Lupus, $\sigma$~Orionis, and Upper Sco are likely drawn from the same parent population \citep{Barenfeld2016,Ansdell2016,Ansdell2017}. To confirm that the $M_{\rm dust}$ populations in $\lambda$~Orionis and Upper Sco are indeed statistically indistinguishable, we again use two-sample tests to compare their stellar mass populations. Using the $M_{\star}$ values from Table~\ref{tab-main} for $\lambda$~Orionis and those from \cite{Barenfeld2016} for Upper Sco, we find $p$-values of 0.77 and 0.11 for the T-test and Wilcoxon rank-sum test, respectively (calculated with {\tt scipy.stats} in Python). Thus the $M_{\star}$ populations are likely drawn from the same parent population and so the $M_{\rm dust}$ distributions can be reliably compared.

\begin{figure}
\begin{center}
\includegraphics[width=8.5cm]{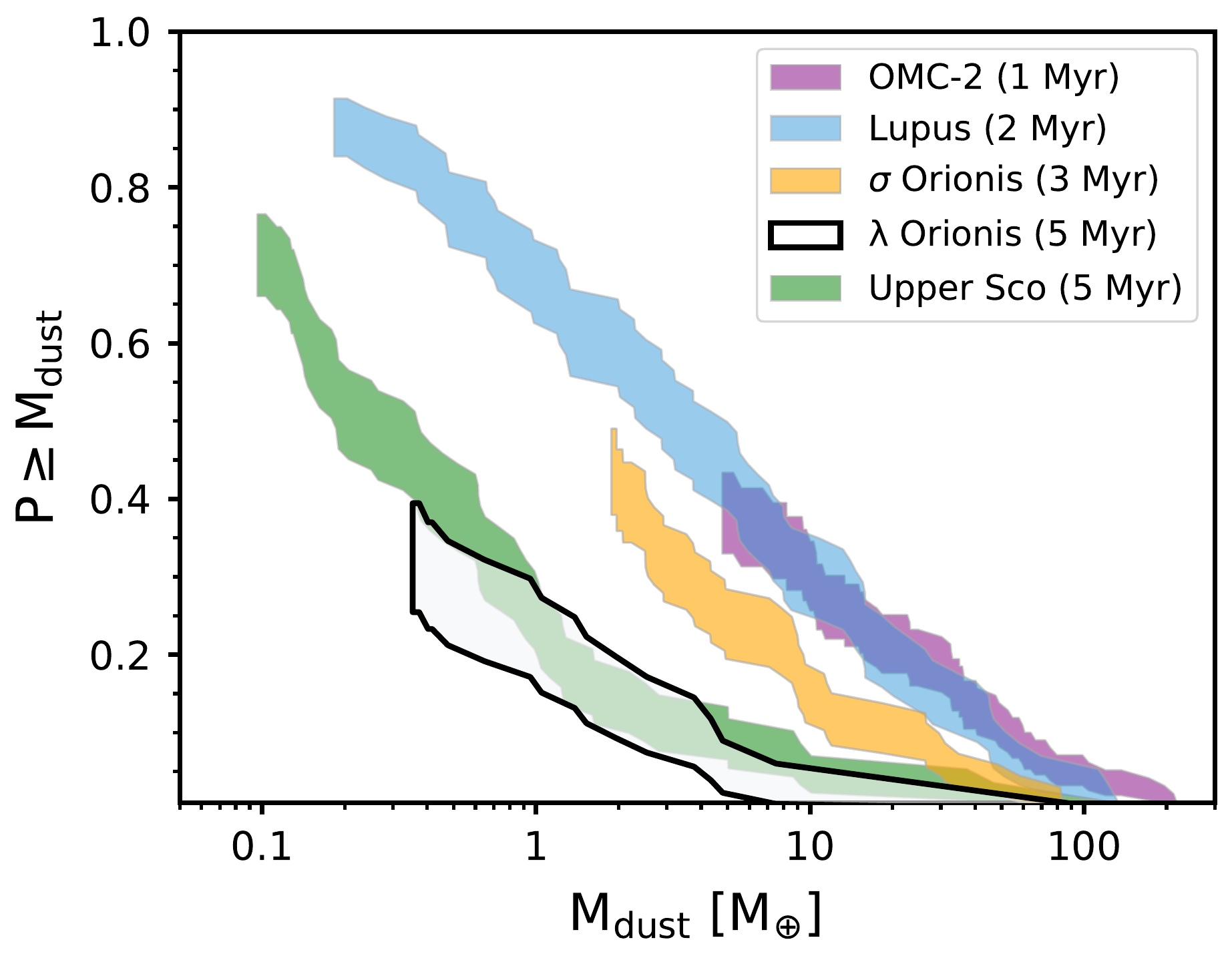}
\caption{The disk dust mass ($M_{\rm dust}$) cumulative distribution for $\lambda$~Orionis compared to several other regions surveyed by ALMA, calculated with the Kaplan-Meier Estimator to account for upper limits (see Section~\ref{sec:compare}). Approximate ages for each region are provided for reference.}
\label{fig:kme}
\end{center}
\end{figure}

Figure~\ref{fig:mdust} also illustrates the decline in $M_{\rm dust}$ distributions with age, but highlights differences at the high-mass end, which are not readily apparent from the cumulative distributions in Figure~\ref{fig:kme}. In Figure~\ref{fig:mdust} , the $\lambda$~Orionis detections are plotted against the upper 32\% of the $M_{\rm dust}$ populations in the comparison regions (32\% was chosen to match the detection rate in $\lambda$~Orionis). Interestingly, the highest-mass disks in all the regions appear to converge around $\sim$100~M$_{\oplus}$ regardless of age or environment. Although the number of disks at such high masses falls steeply after a few Myr, one or more ``outlier" disks appear to persist even at $\sim$5~Myr of age. We discuss these lingering massive disks in Section~\ref{sec:outliers}.

One caveat is that the ages assumed in this work are just the typical median values of the regions reported in the literature. They do not reflect the possibility of distributed age populations, and have been calculated using different methods, any of which are likely imprecise and/or inaccurate due to biases in both the observations and the stellar evolution models \cite[e.g.,][]{Bell2013}. However, the relative ages should be generally reliable and sufficient for the purposes of this work.

\subsection{HD 245185} 
\label{sec:outlier}

HD~245185 hosts a clear outlier disk in $\lambda$~Orionis, having over an order of magnitude higher millimeter continuum and $^{12}$CO flux than the rest of the sample (Figure~\ref{fig:gasdist}). HD~245185 is a well-studied Herbig Ae/Be star in the literature and is the only early-type (A0) star in our sample (Table~\ref{tab-main}) with a mass of $\sim$2.5~$M_\odot$ \cite[e.g.,][]{Folsom2012}. Figure~\ref{fig:gasdet} shows the 1.25~mm continuum map and the $^{12}$CO zero-moment and first-moment maps from our ALMA observations. We note that the proper motions ($\mu_\alpha=0.34$, $\mu_\delta=-1.93$) and distance ($427^{+21}_{-19}$~pc) from {\it Gaia} DR2, as well as the systemic velocity derived from our ALMA $^{12}$CO observations ($v_{\rm LSR}\approx13$~km~s$^{-1}$), are all consistent with cluster membership.

Equation~\ref{eqn-mass} yields $M_{\rm dust}\approx140~M_\oplus$, an order of magnitude higher than the next most massive disk in $\lambda$~Orionis (Figure~\ref{fig:mdust}). One concern is that this is due to HD~245185 being the only hot star in our sample, given that the simplified relation in Equation~\ref{eqn-mass} assumes $T_{\rm dust}=20$~K for all disks. However, only some of the difference (a factor of $\sim$3) may be accounted for if, rather than assuming an isothermal disk, we scale the dust temperature with stellar luminosity using $T_{\rm dust}=25~{\rm K} \times (L_{\star}/L_{\odot})^{0.25}$ as suggested by the radiative transfer model grid of \cite{Andrews2013}. We do not use this scaling, however, as it remains uncertain whether such a clear relationship holds in the real disk population. For example, \cite{Tazzari2017} found no relation between $T_{\rm dust}$ and stellar properties when modeling the ALMA visibilities of resolved Lupus disks. 

Another reason why a massive disk around HD~245185 is unusual at the evolved age of $\lambda$~Orionis is that disks around intermediate-mass stars typically dissipate twice as fast as those around late-type stars \cite[at least based on infrared emission, which traces the warm inner disk; e.g.,][]{Ribas2015}. It is unlikely that HD~245185 is much younger than the average disk in the cluster: several authors have estimated the age of HD 245185, e.g., $6.9\pm2.5$~Myr \citep{Alecian2013} and $5.5\pm2.0$~Myr \citep{Folsom2012}, suggesting we cannot use delayed star formation relative to the rest of $\lambda$~Orionis to reconcile this system. We discuss possible explanations for such long-lived primordial disks in Section~\ref{sec:outliers}.


\section{Discussion} 
\label{sec:discussion}

\subsection{Do Supernovae Impact Planet Formation?} 
\label{sec:sn}

$\lambda$~Orionis may provide a rare snapshot of an evolved SFR that is $\sim$1~Myr post-supernova with the remaining OB stars, lower-mass stellar population, and remnant molecular cloud all still present (Section~\ref{sec:history}). Comparing the disk population in $\lambda$~Orionis to those in other SFRs may therefore provide an opportunity to study how supernovae affect planet formation.

Supernovae are theorized to strip significant amounts of mass from disks around nearby pre-main sequence stars. \cite{CP2017} ran three-dimensional hydrodynamic simulations of protoplanetary disks, with a range of masses and inclinations, subject to a supernova occurring 0.3~pc away. They reported an ``instantaneous stripping" phase with mass-loss rates of 10$^{-5}$~$M_\odot$~yr$^{-1}$ lasting 10--100~years, followed by more moderate but extended ablation with mass-loss rates of 10$^{-6}$ to 10$^{-7}$~$M_\odot$~yr$^{-1}$. For the low-mass (0.1~$M_{\rm Jup}$) and moderate-mass (1.0~$M_{\rm Jup}$) disks in their simulations, up to 90\% and 30\% of the disk mass, respectively, was removed during the instantaneous stripping phase; these disk masses are typical in $\sigma$~Orionis, a possible example of a pre-supernova OB cluster. High-mass (10~$M_{\rm Jup}$) disks, however---similar to the outlier around HD~245185 in $\lambda$~Orionis (Section~\ref{sec:outlier})---were largely unaffected. Since the peak ram pressure in the simulations of \cite{CP2017} strongly depends on distance from the supernova (dropping off as $d^{-3}$), we would expect most disks within 0.3~pc to be significantly depleted in mass, with those further out relatively unaffected.

Indeed, as shown in Figure~\ref{fig:sep}, we observe but do not detect four disks with projected separations $<0.3$~pc from the cluster core, the presumed supernova location (Section~\ref{sec:history}). Several disks are detected just beyond 0.3~pc, but have disk masses similar to the rest of the population. Due to the natural course of cluster expansion, these projected distances are likely overestimates of the source locations when the supernova occurred. Still, our finding that the overall disk population in $\lambda$~Orionis appears to follow the general decline in disk dust mass with age seen in other SFRs, and is statistically indistinguishable from that of the similarly aged Upper Sco association (Figure~\ref{fig:kme}; Section~\ref{sec:compare}), implies that a supernova occurring several Myr into disk evolution does not significantly reduce the amount of planet-forming solid material that would otherwise be available at this age, except potentially for disks that were within a small fraction of a parsec from the supernova event.

\begin{figure}
\begin{center}
\includegraphics[width=8cm]{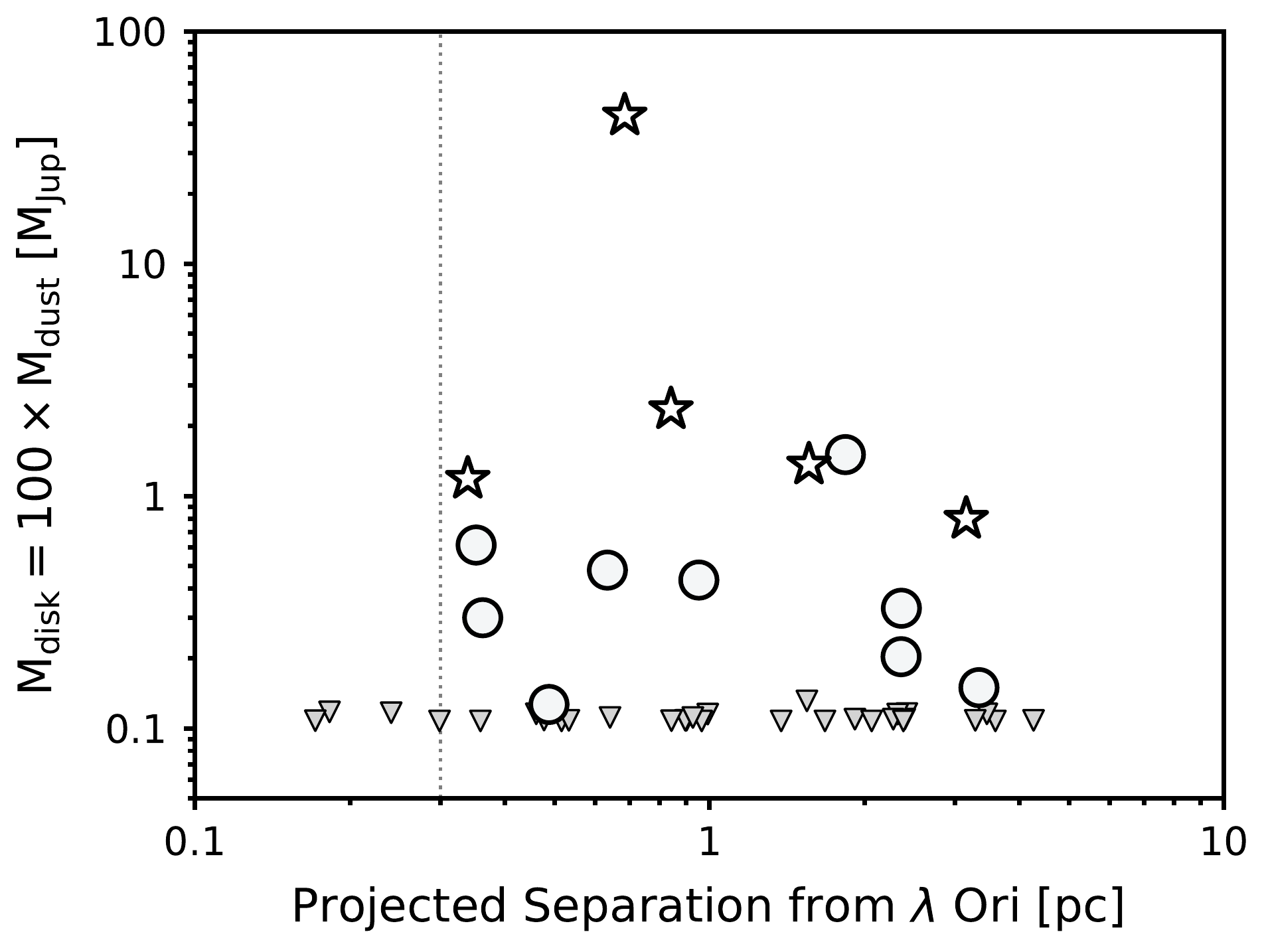}
\caption{Disk mass ($M_{\rm disk}$), assuming a gas-to-dust ratio of 100, as a function of projected separation from the central OB star ($\lambda$~Ori) in $\lambda$~Orionis. Circles are ALMA Band~6 continuum detections and downward-facing triangles are 3$\sigma$ upper limits; stars indicate $^{12}$CO detections. The vertical dotted line denotes the 0.3~pc distance of the supernova in the simulations of \cite{CP2017} (see Section~\ref{sec:sn}).}
\label{fig:sep}
\end{center}
\end{figure}

Additionally, pre-supernova feedback may mute the effects of the actual supernova event on the surrounding environment. Recent simulations that combine stellar winds and photoionization with supernova events \citep{Lucas2020} found that pre-supernova feedback sculpts low-density channels in the gas, through which supernova energy can more freely escape into the wider interstellar medium. As a result, supernova explosions may have only moderate, though more widespread, effects on the surrounding natal molecular clouds and their star/disk populations. When occurring late in disk evolution, it is possible that these moderate effects from supernovae are simply negligible when compared to other disk evolutionary processes.

One caveat to this interpretation, however, is that we cannot rule out that a supernova also occurred in the SFRs against which we are comparing the $\lambda$~Orionis disk population. In particular, a supernova may have also occurred $\sim$1~Myr ago in Upper Sco; this is based on the kinematics of an expanding HI shell in the region \citep{deGeus1992} as well as the kinematic trace-back of the runaway O-type star $\zeta$~Oph and pulsar PSR~1932+11059, which may have once been close binaries in Upper Sco before the pulsar progenitor exploded as a supernova \citep[][]{Hoogerwerf2000,Hoogerwerf2001}. However, the large uncertainties on the present-day kinematics of the runaway objects, including an unknown radial velocity for the pulsar, make the location of the presumed supernova, and thus its potential impact on the Upper Sco disk population, unclear.

Another caveat is that our ALMA sample is selected from targeted {\it Spitzer} observations (Section~\ref{sec:sample}), thus only includes disks that are currently within $\sim$3~pc of the cluster core (see white dashed box in Figure~\ref{fig:pm}). Because the recently identified $\lambda$~Orionis members outside the {\it Spitzer} survey region are radially expanding outward \citep[Figure~\ref{fig:pm};][]{Kounkel2018}, the concern is that our ALMA survey missed some disks that were once much closer to the cluster core and thus potentially most affected by the supernova. However, the faster-moving stars in this radially expanding population only have typical proper motions of $\sim$1~mas~yr$^{-1}$ (relative to the cluster median), which translates to $\sim$2~pc over 1~Myr at 400~pc. Because these stars also have typical projected separations of $\gtrsim$10~pc from the cluster core, it is unlikely that they were particularly close to the supernova when it occurred, especially if the outward acceleration was aided by gravitational feedback of the dispersed gas \citep{Zamora2019}. Moreover, although \cite{Kounkel2018} explain the observed radial motions as due to the ``single-trigger expansion" caused by the supernova, alternative explanations staged pre-supernova are also viable: such kinematics can be a consequence of any mechanism that disperses the intracluster gas \citep[e.g., stellar winds or radiation; see][]{Winter2019}, and pre-supernova feedback may preclude the actual supernova event as the main mechanism driving the removal of the gas potential \citep[e.g.,][]{Lucas2020}. 

 
Indeed, an alternative explanation is that a supernova has not yet actually occurred in $\lambda$~Orionis, and that the observed features described in Section~\ref{sec:history} originate from other aspects of the cluster history, such as pre-supernova feedback. In fact, the expected chemical effects from a supernova are not readily apparent in our current ALMA data. Supernovae are production sites of cosmic rays \cite[see review in][]{GBS2015}, thus the cosmic ray ionization rate of H$_2$ ($\zeta_{\rm CR}$) should be enhanced in $\lambda$~Orionis. Typical ionization rates in molecular clouds are $\zeta_{\rm CR}\sim10^{-17}$~s$^{-1}$ and may be even lower in disk midplanes \citep{Cleeves2014}, however after a supernova the levels can be enhanced to $\zeta_{\rm CR}\sim10^{-15}$ or $10^{-14}$~s$^{-1}$ \cite[e.g.,][]{Indriolo2010,LePetit2016}. At these levels, the transformation of CO into methanol and hydrocarbons proceeds much faster in the ice and gas, on scales of $<$1~Myr rather than 5--10~Myr \citep{Bosman2018,Schwarz2018}. Yet we still detect CO in some $\lambda$~Orionis disks, and at levels expected from their millimeter continuum emission (Figure~\ref{fig:gasdist}). However, the sample size of gas detections is small; deeper observations and additional molecular lines will help determine if our current CO non-detections are due to dispersal of the gas or chemical transformation of the CO. 

Nevertheless, the massive stars in $\lambda$~Orionis, through pre-supernova feedback and/or a recent supernova event itself, appear to have sculpted many of the observational features of the region, yet have not significantly reduced the available planet-forming material in the overall disk population beyond what is expected at this evolved age.

\subsection{External Photoevaporation in $\lambda$ Orionis} 
\label{sec:ep}

OB stars can also impact the disk population through external photoevaporation \citep[e.g.,][]{Johnstone1998, SH1999}, a process that is now thought to be one of the main environmental factors depleting disk material \citep[e.g.,][]{SC2001,Sellek2020}, even in typical galactic UV environments \cite[e.g.,][]{Facchini2016, Haworth2016, Winter2018}. While direct detection of externally driven photoevaporative winds is observationally challenging \citep{Henney1999,Rigliaco2009,Haworth2020}, indirect evidence based on the expected impacts on more easily observable disk properties is growing \citep[e.g.,][]{Fang2012,Mann2014,Kim2016,Guarcello2016,Haworth2017,vanTerwisga2019}. In particular, using ALMA to estimate disk masses, \cite{Ansdell2017} found in $\sigma$~Orionis a dearth of massive ($\gtrsim1~M_{\rm Jup}$) disks within $\sim$0.5~pc of the central OB stars, followed by a clear distant-dependent trend in disk mass out to the cluster edge (see their Figure~6).

Although we do not see a similar distance-dependent trend for the disk masses in $\lambda$~Orionis, there is a lack of even low-mass ($\gtrsim$0.1~$M_{\rm Jup}$) disks within $\sim$0.3~pc of the central OB stars (Figure~\ref{fig:sep}) that could be a tentative signature of external photoevaporation (and/or ablation from the supernova; see Section~\ref{sec:sn}). The lack of a distance-dependent trend in disk mass beyond 0.3~pc may be explained by the older age of the cluster, which has allowed for both more cluster expansion as well as dust grain growth. In this scenario, the sources in our sample were likely closer to the OB stars in the past, experiencing more effective external photoevaporation that removed gas and any entrained small dust grains in the outer disk. As the cluster dispersed with age, the remaining dust grains in the disks would have grown beyond 10--100~$\mu$m, at which point they are too large to be affected by external photoevaporation and instead experience rapid inward radial drift \citep{Haworth2018,Sellek2020}. Thus any detectable distance-dependent trend in disk mass may have been washed out due to the dynamical evolution of the cluster and the evolution of the disk dust population probed by ALMA for a couple Myr beyond the age of $\sigma$~Orionis.

\subsection{Outlier Disks in Evolved Regions} 
\label{sec:outliers}

The handful of relatively nearby SFRs (at $\lesssim$400~pc; the distance of Orion or closer) with evolved ages ($\sim$5--10~Myr old; the end of the disk lifetime) includes $\lambda$~Orionis, Upper Sco, and the TW~Hya Association (TWA). Each of these evolved regions contains at least one long-lived primordial disk that remains much more massive than the rest of the surviving disk population. Younger regions, in contrast, exhibit a smooth distribution in disk masses, regardless of whether they are in low-mass (e.g., Lupus) or high-mass (e.g., OMC-2) SFRs. This is illustrated in Figure~\ref{fig:mdust}, which shows the upper 32\% of the disk dust mass distributions in different SFRs (32\% was chosen to match the detection rate in our $\lambda$~Orionis survey). In Figure~\ref{fig:mdust}, all of the SFRs exhibit similar slopes until the highest-mass disks, at which point the distributions of the evolved Upper Sco and $\lambda$~Orionis regions rise steeply due to their outlier disks (we do not show TWA in Figure~\ref{fig:mdust}, as ALMA surveys of its disk population are not yet published). Additionally, the most massive disks in each SFR, regardless of age or environment, all have $M_{\rm dust}\sim100~M_\oplus$, which suggests that the outlier disks in the evolved regions were {\em not} distinct from birth, but rather that some mechanism stopped the otherwise natural removal of disk material (e.g., by accretion onto the star and/or winds) as the disks evolved. 

A potential explanation is that one or more \mbox{(sub-)stellar} companions formed in these disks, on orbits that are inhibiting the dispersal of outer disk material while also avoiding disk truncation/disruption, resulting in particularly long-lived primordial disks. A single, massive super-Jupiter on an initially circular orbit may excite its own eccentricity \citep[e.g.,][]{DAngelo2006}, enabling the clearing of a large inner cavity while preserving the outer disk; this has been demonstrated for the $\sim$5~Myr old PDS~70 system \citep{Muley2019}, which hosts a super-Jupiter planet at 22~au, and could potentially explain many other transition disks (\citealt{vdMarel2018}; van der Marel et al., subm.). Chains of gas giants \citep[e.g.,][]{Zhu2011} and potentially super-Earths \citep{FC2017,Rosenthal2020} may also open inner disk cavities or large gaps. Although sufficiently inhibiting the accretion of gas-rich disk material (onto either the star or planets) to prolong disk lifetimes remains a challenge in these scenarios, dust evolution models show that giant planets drive strong pressure bumps that effectively prevent millimeter dust from drifting inward; these particle traps then result in prolonged lifetimes for the millimeter dust disks observed by ALMA, especially for higher-mass disks around higher-mass stars \citep[e.g.,][]{Pinilla2020}. Stellar binaries can exert torques on disks that similarly open central cavities, while also effectively inhibiting accretion of material onto the binary. \cite{Alexander2012} showed that binaries on sufficiently close ($\lesssim$1~au) orbits form cavities smaller than the critical radius beyond which photoevaporative winds are efficient, resulting in circumbinary disk lifetimes that are several times longer than those of single-star disks. With these constraints in mind, we discuss below how the outlier disks in $\lambda$~Orionis, Upper Sco, and TWA show evidence of such companions that could explain their extended primordial disk lifetimes.

For HD~245185 in $\lambda$~Orionis, \cite{Kama2015} placed it in the population of Herbig~Ae/Be stars whose stellar abundances are depleted in refractory elements while hosting warm/flared transition disks, as opposed to the population with solar abundances hosting cold/flat full disks. They explained the chemical peculiarity of the former population with giant planets filtering out dusty material as it flows through the disk, resulting in the accretion of high gas-to-dust ratio material onto the star. These massive planets would then also be responsible for clearing out the large gap or cavity in the disk dust structure that defines the transition disk classification \citep{Espaillat2014}. Although our ALMA observations do not resolve any gaps or cavities in the disk around HD~245185 (Figure~\ref{fig:gasdet}), the resolution is poor ($\sim$60~au in radius). As the SED of HD~245185 exhibits a mid-infrared dip \citep{Ansdell2015}, future ALMA observations may still resolve structure in the disk.

For Upper Sco, one of the most massive disks is around 2MASS~J16042165-2130284 (also known simply as J1604), a negligible accretor \citep{Manara2020} of K2 spectral type that hosts a face-on transition disk with a large inner cavity seen in millimeter emission \citep{Mathews2012, Ansdell2020}. Evidence points to one or more high-mass planetary companions clearing out the cavity as well as misaligning an unseen inner disk component that casts variable shadows on the outer disk detected in scattered light \cite[e.g.,][]{Takami2014,Pinilla2018}. This misaligned inner disk is also thought to cause the ``dipper" variability observed in space-based light curves \cite[e.g.,][]{CH2018, Ansdell2020}. Another outlier disk in Upper Sco is around the G-type star 2MASS~J15583692-2257153 (or HD~143006), which also hosts a face-on transition disk exhibiting ``dipper" variability \citep{Ansdell2020}. This disk has a large gap, several narrow rings/gaps, and a bright asymmetry resolved by ALMA \citep{DSHARP2018} suggesting a warped inner disk driven by a low-mass stellar or high-mass planetary companion, potentially orbiting at a few au \citep{Perez2018}. Finally, the most massive disk in Upper Sco is around 2MASS J16113134-1838259 (or AS~205); this is not a transition disk, but displays clear spirals arms detected by ALMA, indicative of strong dynamical interactions induced by a known external companion \citep{Kurtovic2018} that may be dominating its disk structure and evolution.

The outlier disk in TWA is around TW Hya itself. It also has an inner disk warp, this time detected in the gas kinematics \citep{Rosenfeld2012}, which could originate from a massive planetary companion orbiting this K-type star \cite[e.g.,][]{Facchini2014}. Extremely high-resolution ALMA observations have detected a gap at just 1~au, providing further evidence for interactions between the disk and a young planet \citep{Andrews2016}. Moreover, scattered light images taken over nearly two decades show that an azimuthal asymmetry in the outer disk changes in position angle in a manner consistent with shadowing by a precessing warped inner disk due to perturbations by a roughly Jupiter-mass companion orbiting at 1~au \citep{Debes2017}. 

If some or all of these outlier disks are indeed due to sub-stellar companions, they suggest that planet formation, given certain orbital parameters and mass ranges, can change the course of disk evolution to extend the lifetimes of primordial outer disks. More detailed analysis of these and other outlier disks---such as those in the older Upper Cen and Lower Cen regions (e.g., HD~142527, HD~135344B, HD~100453, HD~100546)---could therefore provide insight into the timing and nature of planet formation. For now, we can only make some initial speculations. Early giant planet formation (occurring at $\lesssim$1~Myr) could help explain the outlier disks as well as the apparent discrepancy between the amount of mass in protoplanetary disks compared to mature exoplanet systems  \citep{GR2010,Willialms2012,NK2014,Manara2018b}. We note that observing one or two outlier disks within a given SFR, which typically hosts $\sim$50--100 protoplanetary disks, is broadly consistent with the low occurrence rate of giant planets \cite[e.g.,][]{Cumming2008,Bowler2016,Nielsen2019}, given that particularly massive or multiple gas giants and/or certain orbital parameters are likely required to create outlier disks. If the outlier disks are instead associated with the much more common compact multi-planet systems \cite[e.g.,][]{WF2015}, this could suggest that such systems usually form in the outer disk and migrate inward, during which they disperse the outer disk; only in rarer cases do they form {\it in situ}, preserving the outer regions and resulting in an outlier disk.


\section{Summary} 
\label{sec:summary}

We present an ALMA Band~6 survey of the protoplanetary disk population in the $\lambda$~Orionis OB cluster. This region is important for studying disk evolution and planet formation due to its evolved age and the recent supernova that may have occurred in its core. Our key findings are as follows:

\begin{itemize}

  \item The millimeter emission from $\lambda$~Orionis disks is weak, but not particularly unusual given the cluster's evolved age of $\sim$5~Myr. Only 14 of the 44 disks in our sample are detected in our observations of the 1.25~mm continuum (Figure~\ref{fig:stamps}), which has a 34~$\mu$Jy median rms corresponding to 3$\sigma$ dust mass upper limits of $\sim$0.4~$M_\oplus$. Stacking the 30 non-detections gives a 4$\sigma$ mean signal of 20~$\mu$Jy ($\sim$0.08~$M_\oplus$), indicating that deeper observations should produce more detections. Only 5 disks are also detected in the $^{12}$CO line (Figure~\ref{fig:gasdet}); however, the lack of gas detections is consistent with the weak continuum emission, based on the correlation between millimeter continuum and $^{12}$CO emission seen in younger regions (Figure~\ref{fig:gasdist}).

  \item The effects of massive stars, in the form of pre-supernova feedback and/or a supernova event itself, do not appear to significantly reduce the overall planet-forming capacity of a population of protoplanetary disks that is already a few Myr into evolution. This is based on comparing the disk mass distribution in $\lambda$~Orionis to that of other SFRs, in particular the similarly aged Upper Sco association (Figures~\ref{fig:mdust} and \ref{fig:kme}). One explanation is that supernovae are only effective at stripping mass from nearby disks that are within a small fraction of a parsec. Additionally, pre-supernova feedback may sculpt low-density channels in the intercluster gas, through which energy can more easily escape, significantly muting the impact of supernovae events on the surrounding disk population. However, more work is needed to confirm the occurrence of the supernova event in $\lambda$~Orionis and/or determine whether a recent supernova also occurred in Upper Sco. 

  \item Massive ``outlier" disks lingering in evolved ($\sim$5--10~Myr) regions like $\lambda$~Orionis, Upper Sco, and TWA show evidence for one or more (sub-)stellar companions. Because these massive disks would not be considered outliers in younger ($\sim$1--2~Myr) regions (Figure~\ref{fig:mdust}), their existence suggests that companion formation, including planet formation within certain orbital and mass constraints, can change the course of disk evolution to extend the lifetimes of primordial outer disks. Further study of outlier disks as a population may therefore provide needed insight into the nature and timing of certain types of planet formation.
  
\end{itemize}

Many avenues for future work exist. Deeper ALMA observations of the disk population in $\lambda$~Orionis will build up larger numbers of continuum and line detections to improve our constraints on population-level statistics. Understanding the disk population of the cluster members outside of the {\it Spitzer} survey area will also ensure that the ALMA survey is complete. Obtaining better constraints on the stellar and accretion properties of the disk-hosting stars in $\lambda$~Orionis with wide-band and/or high-resolution spectra will allow us to search for trends seen in other regions, as well as any evidence for external photoevapoaration and viscous evolution. Deeper observations with ALMA of multiple molecular lines tracing the gas content and chemistry will help determine whether our observations reflect disk gas dispersal or the transformation of CO due to enhanced cosmic ray ionization from the supernova. Detailed theoretical studies may also provide insight into the peculiar kinematics of $\lambda$~Orionis and its links to the star-formation history of the region; similar studies in Upper Sco are also important, in particular to assess the possibility of a recent supernova having also occurred in that region. Indeed, $\lambda$~Orionis still has  much to teach us about perhaps one of the most common types of planet-forming environments in the nearby Galaxy.


\acknowledgments

MA and EC acknowledge support from NASA grant NNH18ZDA001N/EW.
TJH is funded by a Royal Society Dorothy Hodgkin Fellowship.
This project has received funding from the European Union's Horizon 2020 research and innovation programme under the Marie Sklodowska-Curie grant agreement No 823823 (DUSTBUSTERS). 
This work was partly supported by the Deutsche Forschungs-Gemeinschaft (DFG, German Research Foundation) - Ref no. FOR 2634/1 TE 1024/1-1.
This work makes use of the following ALMA data: ADS/JAO.ALMA\#2017.1.00466.S and \#2016.1.00447.S. ALMA is a partnership of ESO (representing its member states), NSF (USA) and NINS (Japan), together with NRC (Canada), MOST and ASIAA (Taiwan), and KASI (Republic of Korea), in cooperation with the Republic of Chile. The Joint ALMA Observatory is operated by ESO, AUI/NRAO and NAOJ.
This work made use of data from the European Space Agency (ESA) mission Gaia (\url{https://www.cosmos.esa.int/gaia}), processed by the Gaia Data Processing and Analysis Consortium (DPAC, \url{https://www.cosmos.esa.int/web/gaia/dpac/consortium}). Funding for the DPAC has been provided by national institutions, in particular the institutions participating in the Gaia Multilateral Agreement. 


\vspace{5mm}
\facilities{ALMA, {\it Gaia}}


\software{ {\tt astropy} \citep{AstroPy2013,AstroPy2018}, {\tt ASURV} \citep{Lavalley1992}, {\tt CASA} \citep{CASA2007}, {\tt matplotlib} \citep{Hunter2007}}

\clearpage 



\appendix

\section{ALMA observing log}
\label{app:obslog}

Table~\ref{tab-obs} summarizes the 13 execution blocks that collected the ALMA observations of our $\lambda$~Orionis disk sample (Section~\ref{sec:obs}). It provides the date and time of the observations in Coordinated Universal Time (UTC Date), number of antennae used ($N_{\rm ant}$) and their baseline range ($L_{\rm base}$), precipitable water vapor (PWV) at the time of the observations, and names of the bandpass, flux, and gain calibrators applied in the pipeline calibration.

\begin{deluxetable}{lcccc}[htp]
\tabletypesize{\footnotesize{}} 
\centering 
\tablewidth{500pt} 
\tablecaption{ALMA Observing Log \label{tab-obs}} 
\tablecolumns{5}  
\tablehead{ 
 \colhead{UTC Date} 
&\colhead{$N_{\rm ant}$} 
&\colhead{$L_{\rm base}$} 
&\colhead{PWV} 
&\colhead{Calibrators} \\ 
 \colhead{(End Time)} 
&\colhead{} 
&\colhead{(m)} 
&\colhead{(mm)} 
&\colhead{(Bandpass/Flux, Phase)} 
} 
\startdata 
2018-09-09  & 44 & 15--1213 & 2.1 & J0423$-$0120, \\
(13:39:36)  &    &          &     & J0532$+$0732  \\
2018-09-09  & 46 & 15--1213 & 1.7 & J0510$+$1800, \\  
(10:58:12)  &    &          &     & J0532$+$0732 \\
2018-09-08  & 43 & 15--784  & 1.3 & J0510$+$1800, \\
(13:09:24)  &    &          &     & J0532$+$0732 \\
2018-09-01  & 45 & 15--784  & 1.7 & J0510$+$1800, \\
(12:28:28)  &    &          &     & J0532$+$0732 \\
2018-08-30  & 46 & 15--784  & 0.8 & J0510$+$1800, \\
(12:15:34)  &    &          &     & J0532$+$0732 \\
2018-01-24  & 44 & 15--1398 & 1.4 & J0510$+$1800, \\
(04:01:52)  &    &          &     & J0532$+$0732 \\
2018-01-23  & 43 & 15--1398 & 0.9 & J0510$+$1800, \\
(03:33:37)  &    &          &     & J0532$+$0732 \\
2018-01-22  & 44 & 15--1398 & 0.7 & J0510$+$1800, \\
(04:09:53)  &    &          &     & J0532$+$0732 \\
2018-01-22  & 44 & 15--1398 & 1.0 & J0510$+$1800, \\
(03:00:38)  &    &          &     & J0532$+$0732 \\
2018-01-21  & 44 & 15--1398 & 1.1 & J0510$+$1800, \\
(04:12:04)  &    &          &     & J0532$+$0732 \\
2018-01-21  & 44 & 15--1398 & 1.5 & J0510$+$1800, \\
(03:03:10)  &    &          &     & J0532$+$0732 \\
2018-01-19  & 45 & 15--1398 & 2.4 & J0510$+$1800, \\
(03:44:18)  &    &          &     & J0532$+$0732 \\
2018-01-18  & 44 & 15--1398 & 1.7 & J0510$+$1800, \\
(03:59:12)  &    &          &     & J0532$+$0732 \\
\enddata 
\end{deluxetable} 

\clearpage

\section{ALMA $^{12}$CO Spectra}
\label{app:co}

Figure~\ref{fig:spec} presents the ALMA Band~6 $^{12}$CO spectra for all 44 sources in our $\lambda$~Orionis disk sample. Extraction of the spectra is described in Sections~\ref{sec:co}.

\begin{figure}[htp]
\begin{center}
\includegraphics[width=17.cm]{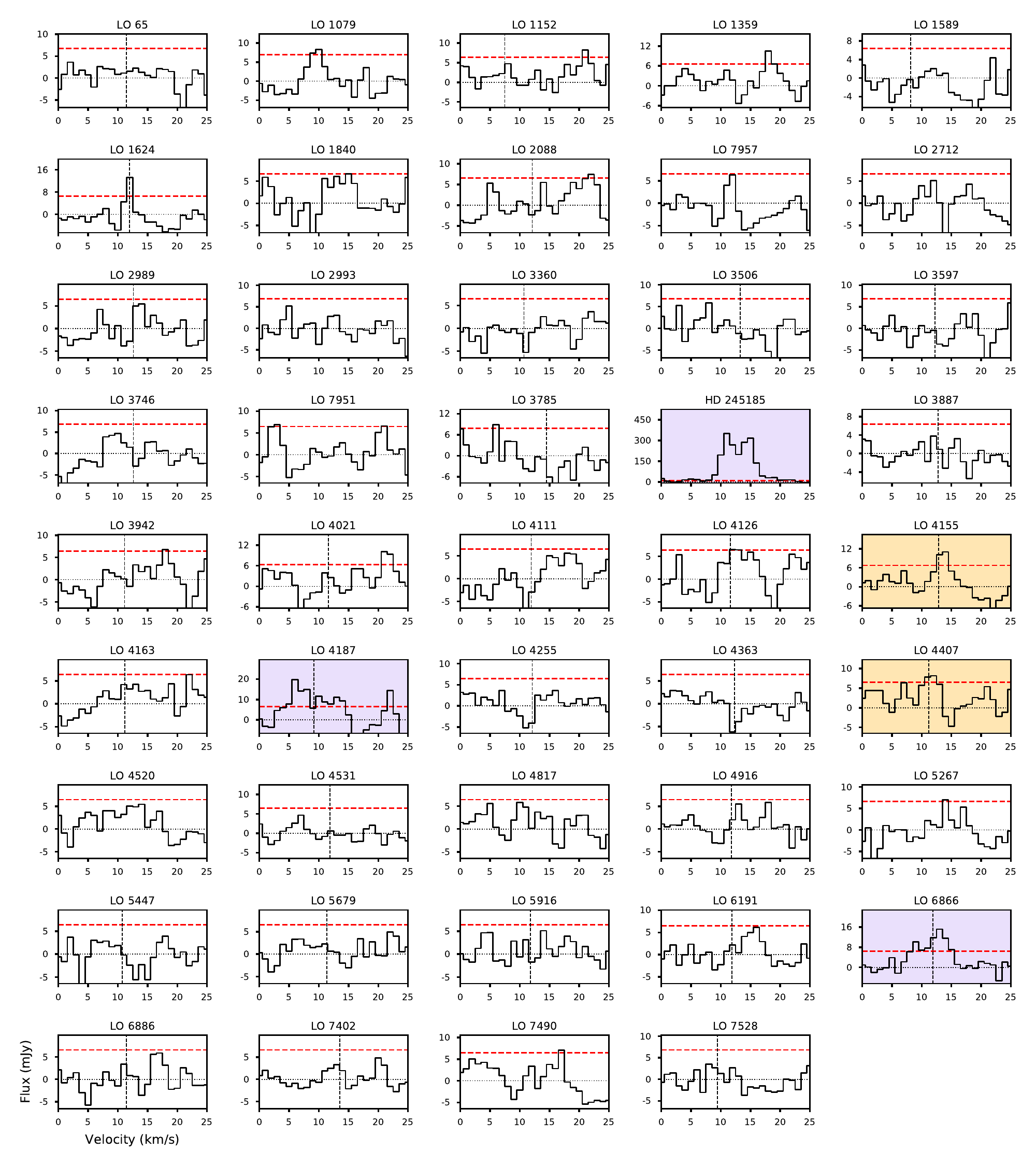}
\caption{ALMA Band~6 $^{12}$CO spectra for $\lambda$~Orionis disks. Horizontal dashed red lines indicate $3\times$ the median channel rms, while horizontal dotted black lines indicate the zero flux level. Vertical dashed lines are the source $v_{\rm LSR}$, when available (Table~\ref{tab-main}). Clear detections ($>4\sigma$) are highlighted in purple, while marginal detections ($\sim$3--4$\sigma$) are indicated with orange.}
\label{fig:spec}
\end{center}
\end{figure}

\clearpage


\begin{thebibliography}{}
\expandafter\ifx\csname natexlab\endcsname\relax\def\natexlab#1{#1}\fi
\providecommand{\url}[1]{\href{#1}{#1}}
\providecommand{\dodoi}[1]{doi:~\href{http://doi.org/#1}{\nolinkurl{#1}}}
\providecommand{\doeprint}[1]{\href{http://ascl.net/#1}{\nolinkurl{http://ascl.net/#1}}}
\providecommand{\doarXiv}[1]{\href{https://arxiv.org/abs/#1}{\nolinkurl{https://arxiv.org/abs/#1}}}

\bibitem[{{Adams}(2010)}]{Adams2010}
{Adams}, F.~C. 2010, \araa, 48, 47, \dodoi{10.1146/annurev-astro-081309-130830}

\bibitem[{{Alecian} {et~al.}(2013){Alecian}, {Wade}, {Catala}, {Grunhut},
  {Landstreet}, {B{\"o}hm}, {Folsom}, \& {Marsden}}]{Alecian2013}
{Alecian}, E., {Wade}, G.~A., {Catala}, C., {et~al.} 2013, \mnras, 429, 1027,
  \dodoi{10.1093/mnras/sts384}

\bibitem[{{Alexander}(2012)}]{Alexander2012}
{Alexander}, R. 2012, \apjl, 757, L29, \dodoi{10.1088/2041-8205/757/2/L29}

\bibitem[{{Andrews}(2015)}]{Andrews2015}
{Andrews}, S.~M. 2015, \pasp, 127, 961, \dodoi{10.1086/683178}

\bibitem[{{Andrews}(2020)}]{Andrews2020}
---. 2020, arXiv e-prints, arXiv:2001.05007.
\newblock \doarXiv{2001.05007}

\bibitem[{{Andrews} {et~al.}(2013){Andrews}, {Rosenfeld}, {Kraus}, \&
  {Wilner}}]{Andrews2013}
{Andrews}, S.~M., {Rosenfeld}, K.~A., {Kraus}, A.~L., \& {Wilner}, D.~J. 2013,
  \apj, 771, 129, \dodoi{10.1088/0004-637X/771/2/129}

\bibitem[{{Andrews} {et~al.}(2018{\natexlab{a}}){Andrews}, {Terrell},
  {Tripathi}, {Ansdell}, {Williams}, \& {Wilner}}]{Andrews2018}
{Andrews}, S.~M., {Terrell}, M., {Tripathi}, A., {et~al.} 2018{\natexlab{a}},
  \apj, 865, 157, \dodoi{10.3847/1538-4357/aadd9f}

\bibitem[{{Andrews} \& {Williams}(2005)}]{Andrews2005}
{Andrews}, S.~M., \& {Williams}, J.~P. 2005, \apj, 631, 1134,
  \dodoi{10.1086/432712}

\bibitem[{{Andrews} {et~al.}(2016){Andrews}, {Wilner}, {Zhu}, {Birnstiel},
  {Carpenter}, {P{\'e}rez}, {Bai}, {{\"O}berg}, {Hughes}, {Isella}, \&
  {Ricci}}]{Andrews2016}
{Andrews}, S.~M., {Wilner}, D.~J., {Zhu}, Z., {et~al.} 2016, \apjl, 820, L40,
  \dodoi{10.3847/2041-8205/820/2/L40}

\bibitem[{{Andrews} {et~al.}(2018{\natexlab{b}}){Andrews}, {Huang},
  {P{\'e}rez}, {Isella}, {Dullemond}, {Kurtovic}, {Guzm{\'a}n}, {Carpenter},
  {Wilner}, {Zhang}, {Zhu}, {Birnstiel}, {Bai}, {Benisty}, {Hughes},
  {{\"O}berg}, \& {Ricci}}]{DSHARP2018}
{Andrews}, S.~M., {Huang}, J., {P{\'e}rez}, L.~M., {et~al.} 2018{\natexlab{b}},
  \apjl, 869, L41, \dodoi{10.3847/2041-8213/aaf741}

\bibitem[{{Ansdell} {et~al.}(2015){Ansdell}, {Williams}, \&
  {Cieza}}]{Ansdell2015}
{Ansdell}, M., {Williams}, J.~P., \& {Cieza}, L.~A. 2015, \apj, 806, 221,
  \dodoi{10.1088/0004-637X/806/2/221}

\bibitem[{{Ansdell} {et~al.}(2017){Ansdell}, {Williams}, {Manara}, {Miotello},
  {Facchini}, {van der Marel}, {Testi}, \& {van Dishoeck}}]{Ansdell2017}
{Ansdell}, M., {Williams}, J.~P., {Manara}, C.~F., {et~al.} 2017, \aj, 153,
  240, \dodoi{10.3847/1538-3881/aa69c0}

\bibitem[{{Ansdell} {et~al.}(2016){Ansdell}, {Williams}, {van der Marel},
  {Carpenter}, {Guidi}, {Hogerheijde}, {Mathews}, {Manara}, {Miotello},
  {Natta}, {Oliveira}, {Tazzari}, {Testi}, {van Dishoeck}, \& {van
  Terwisga}}]{Ansdell2016}
{Ansdell}, M., {Williams}, J.~P., {van der Marel}, N., {et~al.} 2016, \apj,
  828, 46, \dodoi{10.3847/0004-637X/828/1/46}

\bibitem[{{Ansdell} {et~al.}(2018){Ansdell}, {Williams}, {Trapman}, {van
  Terwisga}, {Facchini}, {Manara}, {van der Marel}, {Miotello}, {Tazzari},
  {Hogerheijde}, {Guidi}, {Testi}, \& {van Dishoeck}}]{Ansdell2018}
{Ansdell}, M., {Williams}, J.~P., {Trapman}, L., {et~al.} 2018, \apj, 859, 21,
  \dodoi{10.3847/1538-4357/aab890}

\bibitem[{{Ansdell} {et~al.}(2020){Ansdell}, {Gaidos}, {Hedges}, {Tazzari},
  {Kraus}, {Wyatt}, {Kennedy}, {Williams}, {Mann}, {Angelo}, {D{\^u}chene},
  {Mamajek}, {Carpenter}, {Esplin}, \& {Rizzuto}}]{Ansdell2020}
{Ansdell}, M., {Gaidos}, E., {Hedges}, C., {et~al.} 2020, \mnras, 492, 572,
  \dodoi{10.1093/mnras/stz3361}

\bibitem[{{Astropy Collaboration} {et~al.}(2013){Astropy Collaboration},
  {Robitaille}, {Tollerud}, {Greenfield}, {Droettboom}, {Bray}, {Aldcroft},
  {Davis}, {Ginsburg}, {Price-Whelan}, {Kerzendorf}, {Conley}, {Crighton},
  {Barbary}, {Muna}, {Ferguson}, {Grollier}, {Parikh}, {Nair}, {Unther},
  {Deil}, {Woillez}, {Conseil}, {Kramer}, {Turner}, {Singer}, {Fox}, {Weaver},
  {Zabalza}, {Edwards}, {Azalee Bostroem}, {Burke}, {Casey}, {Crawford},
  {Dencheva}, {Ely}, {Jenness}, {Labrie}, {Lim}, {Pierfederici}, {Pontzen},
  {Ptak}, {Refsdal}, {Servillat}, \& {Streicher}}]{AstroPy2013}
{Astropy Collaboration}, {Robitaille}, T.~P., {Tollerud}, E.~J., {et~al.} 2013,
  \aap, 558, A33, \dodoi{10.1051/0004-6361/201322068}

\bibitem[{{Astropy Collaboration} {et~al.}(2018){Astropy Collaboration},
  {Price-Whelan}, {Sip{\H o}cz}, {G{\"u}nther}, {Lim}, {Crawford}, {Conseil},
  {Shupe}, {Craig}, {Dencheva}, {Ginsburg}, {VanderPlas}, {Bradley},
  {P{\'e}rez-Su{\'a}rez}, {de Val-Borro}, {Aldcroft}, {Cruz}, {Robitaille},
  {Tollerud}, {Ardelean}, {Babej}, {Bach}, {Bachetti}, {Bakanov}, {Bamford},
  {Barentsen}, {Barmby}, {Baumbach}, {Berry}, {Biscani}, {Boquien}, {Bostroem},
  {Bouma}, {Brammer}, {Bray}, {Breytenbach}, {Buddelmeijer}, {Burke},
  {Calderone}, {Cano Rodr{\'{\i}}guez}, {Cara}, {Cardoso}, {Cheedella},
  {Copin}, {Corrales}, {Crichton}, {D'Avella}, {Deil}, {Depagne}, {Dietrich},
  {Donath}, {Droettboom}, {Earl}, {Erben}, {Fabbro}, {Ferreira}, {Finethy},
  {Fox}, {Garrison}, {Gibbons}, {Goldstein}, {Gommers}, {Greco}, {Greenfield},
  {Groener}, {Grollier}, {Hagen}, {Hirst}, {Homeier}, {Horton}, {Hosseinzadeh},
  {Hu}, {Hunkeler}, {Ivezi{\'c}}, {Jain}, {Jenness}, {Kanarek}, {Kendrew},
  {Kern}, {Kerzendorf}, {Khvalko}, {King}, {Kirkby}, {Kulkarni}, {Kumar},
  {Lee}, {Lenz}, {Littlefair}, {Ma}, {Macleod}, {Mastropietro}, {McCully},
  {Montagnac}, {Morris}, {Mueller}, {Mumford}, {Muna}, {Murphy}, {Nelson},
  {Nguyen}, {Ninan}, {N{\"o}the}, {Ogaz}, {Oh}, {Parejko}, {Parley}, {Pascual},
  {Patil}, {Patil}, {Plunkett}, {Prochaska}, {Rastogi}, {Reddy Janga},
  {Sabater}, {Sakurikar}, {Seifert}, {Sherbert}, {Sherwood-Taylor}, {Shih},
  {Sick}, {Silbiger}, {Singanamalla}, {Singer}, {Sladen}, {Sooley},
  {Sornarajah}, {Streicher}, {Teuben}, {Thomas}, {Tremblay}, {Turner},
  {Terr{\'o}n}, {van Kerkwijk}, {de la Vega}, {Watkins}, {Weaver}, {Whitmore},
  {Woillez}, {Zabalza}, \& {Astropy Contributors}}]{AstroPy2018}
{Astropy Collaboration}, {Price-Whelan}, A.~M., {Sip{\H o}cz}, B.~M., {et~al.}
  2018, \aj, 156, 123, \dodoi{10.3847/1538-3881/aabc4f}

\bibitem[{{Bailer-Jones} {et~al.}(2018){Bailer-Jones}, {Rybizki}, {Fouesneau},
  {Mantelet}, \& {Andrae}}]{BJ2018}
{Bailer-Jones}, C.~A.~L., {Rybizki}, J., {Fouesneau}, M., {Mantelet}, G., \&
  {Andrae}, R. 2018, \aj, 156, 58, \dodoi{10.3847/1538-3881/aacb21}

\bibitem[{{Baraffe} {et~al.}(2015){Baraffe}, {Homeier}, {Allard}, \&
  {Chabrier}}]{Baraffe2015}
{Baraffe}, I., {Homeier}, D., {Allard}, F., \& {Chabrier}, G. 2015, \aap, 577,
  A42, \dodoi{10.1051/0004-6361/201425481}

\bibitem[{{Barenfeld} {et~al.}(2016){Barenfeld}, {Carpenter}, {Ricci}, \&
  {Isella}}]{Barenfeld2016}
{Barenfeld}, S.~A., {Carpenter}, J.~M., {Ricci}, L., \& {Isella}, A. 2016,
  \apj, 827, 142, \dodoi{10.3847/0004-637X/827/2/142}

\bibitem[{{Bayo} {et~al.}(2011){Bayo}, {Barrado}, {Stauffer},
  {Morales-Calder{\'o}n}, {Melo}, {Hu{\'e}lamo}, {Bouy}, {Stelzer}, {Tamura},
  \& {Jayawardhana}}]{Bayo2011}
{Bayo}, A., {Barrado}, D., {Stauffer}, J., {et~al.} 2011, \aap, 536, A63,
  \dodoi{10.1051/0004-6361/201116617}

\bibitem[{{Beckwith} {et~al.}(1990){Beckwith}, {Sargent}, {Chini}, \&
  {Guesten}}]{Beckwith1990}
{Beckwith}, S. V.~W., {Sargent}, A.~I., {Chini}, R.~S., \& {Guesten}, R. 1990,
  \aj, 99, 924, \dodoi{10.1086/115385}

\bibitem[{{Bell} {et~al.}(2013){Bell}, {Naylor}, {Mayne}, {Jeffries}, \&
  {Littlefair}}]{Bell2013}
{Bell}, C. P.~M., {Naylor}, T., {Mayne}, N.~J., {Jeffries}, R.~D., \&
  {Littlefair}, S.~P. 2013, \mnras, 434, 806, \dodoi{10.1093/mnras/stt1075}

\bibitem[{{Booth} \& {Ilee}(2020)}]{Booth2020}
{Booth}, A.~S., \& {Ilee}, J.~D. 2020, \mnras, 493, L108,
  \dodoi{10.1093/mnrasl/slaa014}

\bibitem[{{Bosman} {et~al.}(2018){Bosman}, {Walsh}, \& {van
  Dishoeck}}]{Bosman2018}
{Bosman}, A.~D., {Walsh}, C., \& {van Dishoeck}, E.~F. 2018, \aap, 618, A182,
  \dodoi{10.1051/0004-6361/201833497}

\bibitem[{{Bowler}(2016)}]{Bowler2016}
{Bowler}, B.~P. 2016, \pasp, 128, 102001,
  \dodoi{10.1088/1538-3873/128/968/102001}

\bibitem[{{Cazzoletti} {et~al.}(2019){Cazzoletti}, {Manara}, {Baobab Liu}, {van
  Dishoeck}, {Facchini}, {Alcal{\`a}}, {Ansdell}, {Testi}, {Williams},
  {Carrasco-Gonz{\'a}lez}, {Dong}, {Forbrich}, {Fukagawa}, {Galv{\'a}n-Madrid},
  {Hirano}, {Hogerheijde}, {Hasegawa}, {Muto}, {Pinilla}, {Takami}, {Tamura},
  {Tazzari}, \& {Wisniewski}}]{Cazzoletti2019}
{Cazzoletti}, P., {Manara}, C.~F., {Baobab Liu}, H., {et~al.} 2019, \aap, 626,
  A11, \dodoi{10.1051/0004-6361/201935273}

\bibitem[{{Cieza} {et~al.}(2019){Cieza}, {Ru{\'\i}z-Rodr{\'\i}guez}, {Hales},
  {Casassus}, {P{\'e}rez}, {Gonzalez-Ruilova}, {C{\'a}novas}, {Williams},
  {Zurlo}, {Ansdell}, {Avenhaus}, {Bayo}, {Bertrang}, {Christiaens}, {Dent},
  {Ferrero}, {Gamen}, {Olofsson}, {Orcajo}, {Pe{\~n}a Ram{\'\i}rez},
  {Principe}, {Schreiber}, \& {van der Plas}}]{Cieza2019}
{Cieza}, L.~A., {Ru{\'\i}z-Rodr{\'\i}guez}, D., {Hales}, A., {et~al.} 2019,
  \mnras, 482, 698, \dodoi{10.1093/mnras/sty2653}

\bibitem[{{Clarke}(2007)}]{Clarke2007}
{Clarke}, C.~J. 2007, \mnras, 376, 1350,
  \dodoi{10.1111/j.1365-2966.2007.11547.x}

\bibitem[{{Cleeves} {et~al.}(2014){Cleeves}, {Bergin}, \&
  {Adams}}]{Cleeves2014}
{Cleeves}, L.~I., {Bergin}, E.~A., \& {Adams}, F.~C. 2014, \apj, 794, 123,
  \dodoi{10.1088/0004-637X/794/2/123}

\bibitem[{{Close} \& {Pittard}(2017)}]{CP2017}
{Close}, J.~L., \& {Pittard}, J.~M. 2017, \mnras, 469, 1117,
  \dodoi{10.1093/mnras/stx897}

\bibitem[{{Cody} \& {Hillenbrand}(2018)}]{CH2018}
{Cody}, A.~M., \& {Hillenbrand}, L.~A. 2018, \aj, 156, 71,
  \dodoi{10.3847/1538-3881/aacead}

\bibitem[{{Concha-Ram{\'\i}rez} {et~al.}(2019){Concha-Ram{\'\i}rez}, {Wilhelm},
  {Portegies Zwart}, \& {Haworth}}]{CM2019}
{Concha-Ram{\'\i}rez}, F., {Wilhelm}, M. J.~C., {Portegies Zwart}, S., \&
  {Haworth}, T.~J. 2019, \mnras, 490, 5678, \dodoi{10.1093/mnras/stz2973}

\bibitem[{{Cumming} {et~al.}(2008){Cumming}, {Butler}, {Marcy}, {Vogt},
  {Wright}, \& {Fischer}}]{Cumming2008}
{Cumming}, A., {Butler}, R.~P., {Marcy}, G.~W., {et~al.} 2008, \pasp, 120, 531,
  \dodoi{10.1086/588487}

\bibitem[{{Cunha} \& {Smith}(1996)}]{CS1996}
{Cunha}, K., \& {Smith}, V.~V. 1996, \aap, 309, 892

\bibitem[{{Dale} {et~al.}(2014){Dale}, {Ngoumou}, {Ercolano}, \&
  {Bonnell}}]{Dale2014}
{Dale}, J.~E., {Ngoumou}, J., {Ercolano}, B., \& {Bonnell}, I.~A. 2014, \mnras,
  442, 694, \dodoi{10.1093/mnras/stu816}

\bibitem[{{D'Angelo} {et~al.}(2006){D'Angelo}, {Lubow}, \&
  {Bate}}]{DAngelo2006}
{D'Angelo}, G., {Lubow}, S.~H., \& {Bate}, M.~R. 2006, \apj, 652, 1698,
  \dodoi{10.1086/508451}

\bibitem[{Davidson-Pilon {et~al.}(2020)Davidson-Pilon, Kalderstam, Jacobson,
  sean reed, Kuhn, Zivich, Williamson, AbdealiJK, Datta, Fiore-Gartland, Parij,
  WIlson, Gabriel, Moneda, Stark, Moncada-Torres, Gadgil, Jona, Singaravelan,
  Besson, Peña, Anton, Klintberg, Noorbakhsh, Begun, Kumar, Hussey, Golland,
  jlim13, \& Flaxman}]{lifelines}
Davidson-Pilon, C., Kalderstam, J., Jacobson, N., {et~al.} 2020,
  CamDavidsonPilon/lifelines: v0.24.15, v0.24.15,  Zenodo,
  \dodoi{10.5281/zenodo.3934629}

\bibitem[{{de Geus}(1992)}]{deGeus1992}
{de Geus}, E.~J. 1992, \aap, 262, 258

\bibitem[{{Debes} {et~al.}(2017){Debes}, {Poteet}, {Jang-Condell}, {Gaspar},
  {Hines}, {Kastner}, {Pueyo}, {Rapson}, {Roberge}, {Schneider}, \&
  {Weinberger}}]{Debes2017}
{Debes}, J.~H., {Poteet}, C.~A., {Jang-Condell}, H., {et~al.} 2017, \apj, 835,
  205, \dodoi{10.3847/1538-4357/835/2/205}

\bibitem[{{Diplas} \& {Savage}(1994)}]{DS1994}
{Diplas}, A., \& {Savage}, B.~D. 1994, \apjs, 93, 211, \dodoi{10.1086/192052}

\bibitem[{{Dolan} \& {Mathieu}(1999)}]{DM1999}
{Dolan}, C.~J., \& {Mathieu}, R.~D. 1999, \aj, 118, 2409,
  \dodoi{10.1086/301075}

\bibitem[{{Dolan} \& {Mathieu}(2001)}]{DM2001}
---. 2001, \aj, 121, 2124, \dodoi{10.1086/319946}

\bibitem[{{Eistrup} {et~al.}(2016){Eistrup}, {Walsh}, \& {van
  Dishoeck}}]{Eistrup2016}
{Eistrup}, C., {Walsh}, C., \& {van Dishoeck}, E.~F. 2016, \aap, 595, A83,
  \dodoi{10.1051/0004-6361/201628509}

\bibitem[{{Espaillat} {et~al.}(2014){Espaillat}, {Muzerolle}, {Najita},
  {Andrews}, {Zhu}, {Calvet}, {Kraus}, {Hashimoto}, {Kraus}, \&
  {D'Alessio}}]{Espaillat2014}
{Espaillat}, C., {Muzerolle}, J., {Najita}, J., {et~al.} 2014, in Protostars
  and Planets VI, ed. H.~{Beuther}, R.~S. {Klessen}, C.~P. {Dullemond}, \&
  T.~{Henning}, 497, \dodoi{10.2458/azu_uapress_9780816531240-ch022}

\bibitem[{{Facchini} {et~al.}(2016){Facchini}, {Clarke}, \&
  {Bisbas}}]{Facchini2016}
{Facchini}, S., {Clarke}, C.~J., \& {Bisbas}, T.~G. 2016, \mnras, 457, 3593,
  \dodoi{10.1093/mnras/stw240}

\bibitem[{{Facchini} {et~al.}(2014){Facchini}, {Ricci}, \&
  {Lodato}}]{Facchini2014}
{Facchini}, S., {Ricci}, L., \& {Lodato}, G. 2014, \mnras, 442, 3700,
  \dodoi{10.1093/mnras/stu1149}

\bibitem[{{Fang} {et~al.}(2012){Fang}, {van Boekel}, {King}, {Henning},
  {Bouwman}, {Doi}, {Okamoto}, {Roccatagliata}, \&
  {Sicilia-Aguilar}}]{Fang2012}
{Fang}, M., {van Boekel}, R., {King}, R.~R., {et~al.} 2012, \aap, 539, A119,
  \dodoi{10.1051/0004-6361/201015914}

\bibitem[{{Fazio} {et~al.}(2004){Fazio}, {Hora}, {Allen}, {Ashby}, {Barmby},
  {Deutsch}, {Huang}, {Kleiner}, {Marengo}, {Megeath}, {Melnick}, {Pahre},
  {Patten}, {Polizotti}, {Smith}, {Taylor}, {Wang}, {Willner}, {Hoffmann},
  {Pipher}, {Forrest}, {McMurty}, {McCreight}, {McKelvey}, {McMurray}, {Koch},
  {Moseley}, {Arendt}, {Mentzell}, {Marx}, {Losch}, {Mayman}, {Eichhorn},
  {Krebs}, {Jhabvala}, {Gezari}, {Fixsen}, {Flores}, {Shakoorzadeh}, {Jungo},
  {Hakun}, {Workman}, {Karpati}, {Kichak}, {Whitley}, {Mann}, {Tollestrup},
  {Eisenhardt}, {Stern}, {Gorjian}, {Bhattacharya}, {Carey}, {Nelson},
  {Glaccum}, {Lacy}, {Lowrance}, {Laine}, {Reach}, {Stauffer}, {Surace},
  {Wilson}, {Wright}, {Hoffman}, {Domingo}, \& {Cohen}}]{Fazio2004}
{Fazio}, G.~G., {Hora}, J.~L., {Allen}, L.~E., {et~al.} 2004, \apjs, 154, 10,
  \dodoi{10.1086/422843}

\bibitem[{{Folsom} {et~al.}(2012){Folsom}, {Bagnulo}, {Wade}, {Alecian},
  {Landstreet}, {Marsden}, \& {Waite}}]{Folsom2012}
{Folsom}, C.~P., {Bagnulo}, S., {Wade}, G.~A., {et~al.} 2012, \mnras, 422,
  2072, \dodoi{10.1111/j.1365-2966.2012.20718.x}

\bibitem[{{Fung} \& {Chiang}(2017)}]{FC2017}
{Fung}, J., \& {Chiang}, E. 2017, \apj, 839, 100,
  \dodoi{10.3847/1538-4357/aa6934}

\bibitem[{{Gaia Collaboration} {et~al.}(2016){Gaia Collaboration}, {Prusti},
  {de Bruijne}, {Brown}, {Vallenari}, {Babusiaux}, {Bailer-Jones}, {Bastian},
  {Biermann}, {Evans}, {Eyer}, {Jansen}, {Jordi}, {Klioner}, {Lammers},
  {Lindegren}, {Luri}, {Mignard}, {Milligan}, {Panem}, {Poinsignon},
  {Pourbaix}, {Randich}, {Sarri}, {Sartoretti}, {Siddiqui}, {Soubiran},
  {Valette}, {van Leeuwen}, {Walton}, {Aerts}, {Arenou}, {Cropper}, {Drimmel},
  {H{\o}g}, {Katz}, {Lattanzi}, {O'Mullane}, {Grebel}, {Holland}, {Huc},
  {Passot}, {Bramante}, {Cacciari}, {Casta{\~n}eda}, {Chaoul}, {Cheek}, {De
  Angeli}, {Fabricius}, {Guerra}, {Hern{\'a}ndez}, {Jean-Antoine-Piccolo},
  {Masana}, {Messineo}, {Mowlavi}, {Nienartowicz}, {Ord{\'o}{\~n}ez-Blanco},
  {Panuzzo}, {Portell}, {Richards}, {Riello}, {Seabroke}, {Tanga},
  {Th{\'e}venin}, {Torra}, {Els}, {Gracia-Abril}, {Comoretto},
  {Garcia-Reinaldos}, {Lock}, {Mercier}, {Altmann}, {Andrae}, {Astraatmadja},
  {Bellas-Velidis}, {Benson}, {Berthier}, {Blomme}, {Busso}, {Carry},
  {Cellino}, {Clementini}, {Cowell}, {Creevey}, {Cuypers}, {Davidson}, {De
  Ridder}, {de Torres}, {Delchambre}, {Dell'Oro}, {Ducourant}, {Fr{\'e}mat},
  {Garc{\'\i}a-Torres}, {Gosset}, {Halbwachs}, {Hambly}, {Harrison}, {Hauser},
  {Hestroffer}, {Hodgkin}, {Huckle}, {Hutton}, {Jasniewicz}, {Jordan},
  {Kontizas}, {Korn}, {Lanzafame}, {Manteiga}, {Moitinho}, {Muinonen},
  {Osinde}, {Pancino}, {Pauwels}, {Petit}, {Recio-Blanco}, {Robin}, {Sarro},
  {Siopis}, {Smith}, {Smith}, {Sozzetti}, {Thuillot}, {van Reeven}, {Viala},
  {Abbas}, {Abreu Aramburu}, {Accart}, {Aguado}, {Allan}, {Allasia},
  {Altavilla}, {{\'A}lvarez}, {Alves}, {Anderson}, {Andrei}, {Anglada Varela},
  {Antiche}, {Antoja}, {Ant{\'o}n}, {Arcay}, {Atzei}, {Ayache}, {Bach},
  {Baker}, {Balaguer-N{\'u}{\~n}ez}, {Barache}, {Barata}, {Barbier}, {Barblan},
  {Baroni}, {Barrado y Navascu{\'e}s}, {Barros}, {Barstow}, {Becciani},
  {Bellazzini}, {Bellei}, {Bello Garc{\'\i}a}, {Belokurov}, {Bendjoya},
  {Berihuete}, {Bianchi}, {Bienaym{\'e}}, {Billebaud}, {Blagorodnova},
  {Blanco-Cuaresma}, {Boch}, {Bombrun}, {Borrachero}, {Bouquillon}, {Bourda},
  {Bouy}, {Bragaglia}, {Breddels}, {Brouillet}, {Br{\"u}semeister},
  {Bucciarelli}, {Budnik}, {Burgess}, {Burgon}, {Burlacu}, {Busonero}, {Buzzi},
  {Caffau}, {Cambras}, {Campbell}, {Cancelliere}, {Cantat-Gaudin}, {Carlucci},
  {Carrasco}, {Castellani}, {Charlot}, {Charnas}, {Charvet}, {Chassat},
  {Chiavassa}, {Clotet}, {Cocozza}, {Collins}, {Collins}, {Costigan}, {Crifo},
  {Cross}, {Crosta}, {Crowley}, {Dafonte}, {Damerdji}, {Dapergolas}, {David},
  {David}, {De Cat}, {de Felice}, {de Laverny}, {De Luise}, {De March}, {de
  Martino}, {de Souza}, {Debosscher}, {del Pozo}, {Delbo}, {Delgado},
  {Delgado}, {di Marco}, {Di Matteo}, {Diakite}, {Distefano}, {Dolding}, {Dos
  Anjos}, {Drazinos}, {Dur{\'a}n}, {Dzigan}, {Ecale}, {Edvardsson}, {Enke},
  {Erdmann}, {Escolar}, {Espina}, {Evans}, {Eynard Bontemps}, {Fabre},
  {Fabrizio}, {Faigler}, {Falc{\~a}o}, {Farr{\`a}s Casas}, {Faye}, {Federici},
  {Fedorets}, {Fern{\'a}ndez-Hern{\'a}ndez}, {Fernique}, {Fienga}, {Figueras},
  {Filippi}, {Findeisen}, {Fonti}, {Fouesneau}, {Fraile}, {Fraser}, {Fuchs},
  {Furnell}, {Gai}, {Galleti}, {Galluccio}, {Garabato}, {Garc{\'\i}a-Sedano},
  {Gar{\'e}}, {Garofalo}, {Garralda}, {Gavras}, {Gerssen}, {Geyer}, {Gilmore},
  {Girona}, {Giuffrida}, {Gomes}, {Gonz{\'a}lez-Marcos},
  {Gonz{\'a}lez-N{\'u}{\~n}ez}, {Gonz{\'a}lez-Vidal}, {Granvik}, {Guerrier},
  {Guillout}, {Guiraud}, {G{\'u}rpide}, {Guti{\'e}rrez-S{\'a}nchez}, {Guy},
  {Haigron}, {Hatzidimitriou}, {Haywood}, {Heiter}, {Helmi}, {Hobbs},
  {Hofmann}, {Holl}, {Holland }, {Hunt}, {Hypki}, {Icardi}, {Irwin}, {Jevardat
  de Fombelle}, {Jofr{\'e}}, {Jonker}, {Jorissen}, {Julbe}, {Karampelas},
  {Kochoska}, {Kohley}, {Kolenberg}, {Kontizas}, {Koposov}, {Kordopatis},
  {Koubsky}, {Kowalczyk}, {Krone-Martins}, {Kudryashova}, {Kull}, {Bachchan},
  {Lacoste-Seris}, {Lanza}, {Lavigne}, {Le Poncin-Lafitte}, {Lebreton},
  {Lebzelter}, {Leccia}, {Leclerc}, {Lecoeur-Taibi}, {Lemaitre}, {Lenhardt},
  {Leroux}, {Liao}, {Licata}, {Lindstr{\o}m}, {Lister}, {Livanou}, {Lobel},
  {L{\"o}ffler}, {L{\'o}pez}, {Lopez-Lozano}, {Lorenz}, {Loureiro},
  {MacDonald}, {Magalh{\~a}es Fernandes}, {Managau}, {Mann}, {Mantelet},
  {Marchal}, {Marchant}, {Marconi}, {Marie}, {Marinoni}, {Marrese},
  {Marschalk{\'o}}, {Marshall}, {Mart{\'\i}n-Fleitas}, {Martino}, {Mary},
  {Matijevi{\v{c}}}, {Mazeh}, {McMillan}, {Messina}, {Mestre}, {Michalik},
  {Millar}, {Miranda}, {Molina}, {Molinaro}, {Molinaro}, {Moln{\'a}r},
  {Moniez}, {Montegriffo}, {Monteiro}, {Mor}, {Mora}, {Morbidelli}, {Morel},
  {Morgenthaler}, {Morley}, {Morris}, {Mulone}, {Muraveva}, {Musella},
  {Narbonne}, {Nelemans}, {Nicastro}, {Noval}, {Ord{\'e}novic},
  {Ordieres-Mer{\'e}}, {Osborne}, {Pagani}, {Pagano}, {Pailler}, {Palacin},
  {Palaversa}, {Parsons}, {Paulsen}, {Pecoraro}, {Pedrosa}, {Pentik{\"a}inen},
  {Pereira}, {Pichon}, {Piersimoni}, {Pineau}, {Plachy}, {Plum}, {Poujoulet},
  {Pr{\v{s}}a}, {Pulone}, {Ragaini}, {Rago}, {Rambaux}, {Ramos-Lerate},
  {Ranalli}, {Rauw}, {Read}, {Regibo}, {Renk}, {Reyl{\'e}}, {Ribeiro},
  {Rimoldini}, {Ripepi}, {Riva}, {Rixon}, {Roelens}, {Romero-G{\'o}mez},
  {Rowell}, {Royer}, {Rudolph}, {Ruiz-Dern}, {Sadowski}, {Sagrist{\`a}
  Sell{\'e}s}, {Sahlmann}, {Salgado}, {Salguero}, {Sarasso}, {Savietto},
  {Schnorhk}, {Schultheis}, {Sciacca}, {Segol}, {Segovia}, {Segransan},
  {Serpell}, {Shih}, {Smareglia}, {Smart}, {Smith}, {Solano}, {Solitro},
  {Sordo}, {Soria Nieto}, {Souchay}, {Spagna}, {Spoto}, {Stampa}, {Steele},
  {Steidelm{\"u}ller}, {Stephenson}, {Stoev}, {Suess}, {S{\"u}veges}, {Surdej},
  {Szabados}, {Szegedi-Elek}, {Tapiador}, {Taris}, {Tauran}, {Taylor},
  {Teixeira}, {Terrett}, {Tingley}, {Trager}, {Turon}, {Ulla}, {Utrilla},
  {Valentini}, {van Elteren}, {Van Hemelryck}, {van Leeuwen}, {Varadi},
  {Vecchiato}, {Veljanoski}, {Via}, {Vicente}, {Vogt}, {Voss}, {Votruba},
  {Voutsinas}, {Walmsley}, {Weiler}, {Weingrill}, {Werner}, {Wevers},
  {Whitehead}, {Wyrzykowski}, {Yoldas}, {{\v{Z}}erjal}, {Zucker}, {Zurbach},
  {Zwitter}, {Alecu}, {Allen}, {Allende Prieto}, {Amorim},
  {Anglada-Escud{\'e}}, {Arsenijevic}, {Azaz}, {Balm}, {Beck}, {Bernstein},
  {Bigot}, {Bijaoui}, {Blasco}, {Bonfigli}, {Bono}, {Boudreault}, {Bressan},
  {Brown}, {Brunet}, {Bunclark}, {Buonanno}, {Butkevich}, {Carret}, {Carrion},
  {Chemin}, {Ch{\'e}reau}, {Corcione}, {Darmigny}, {de Boer}, {de Teodoro}, {de
  Zeeuw}, {Delle Luche}, {Domingues}, {Dubath}, {Fodor}, {Fr{\'e}zouls},
  {Fries}, {Fustes}, {Fyfe}, {Gallardo}, {Gallegos}, {Gardiol}, {Gebran},
  {Gomboc}, {G{\'o}mez}, {Grux}, {Gueguen}, {Heyrovsky}, {Hoar}, {Iannicola},
  {Isasi Parache}, {Janotto}, {Joliet}, {Jonckheere}, {Keil}, {Kim},
  {Klagyivik}, {Klar}, {Knude}, {Kochukhov}, {Kolka}, {Kos}, {Kutka}, {Lainey},
  {LeBouquin}, {Liu}, {Loreggia}, {Makarov}, {Marseille}, {Martayan},
  {Martinez-Rubi}, {Massart}, {Meynadier}, {Mignot}, {Munari}, {Nguyen},
  {Nordlander}, {Ocvirk}, {O'Flaherty}, {Olias Sanz}, {Ortiz}, {Osorio},
  {Oszkiewicz}, {Ouzounis}, {Palmer}, {Park}, {Pasquato}, {Peltzer}, {Peralta},
  {P{\'e}turaud}, {Pieniluoma}, {Pigozzi}, {Poels}, {Prat}, {Prod'homme},
  {Raison}, {Rebordao}, {Risquez}, {Rocca-Volmerange}, {Rosen}, {Ruiz-Fuertes},
  {Russo}, {Sembay}, {Serraller Vizcaino}, {Short}, {Siebert}, {Silva},
  {Sinachopoulos}, {Slezak}, {Soffel}, {Sosnowska}, {Strai{\v{z}}ys}, {ter
  Linden}, {Terrell}, {Theil}, {Tiede}, {Troisi}, {Tsalmantza}, {Tur},
  {Vaccari}, {Vachier}, {Valles}, {Van Hamme}, {Veltz}, {Virtanen}, {Wallut},
  {Wichmann}, {Wilkinson}, {Ziaeepour}, \& {Zschocke}}]{Gaia2016}
{Gaia Collaboration}, {Prusti}, T., {de Bruijne}, J.~H.~J., {et~al.} 2016,
  \aap, 595, A1, \dodoi{10.1051/0004-6361/201629272}

\bibitem[{{Gaia Collaboration} {et~al.}(2018){Gaia Collaboration}, {Brown},
  {Vallenari}, {Prusti}, {de Bruijne}, {Babusiaux}, {Bailer-Jones}, {Biermann},
  {Evans}, {Eyer}, {Jansen}, {Jordi}, {Klioner}, {Lammers}, {Lindegren},
  {Luri}, {Mignard}, {Panem}, {Pourbaix}, {Randich}, {Sartoretti}, {Siddiqui},
  {Soubiran}, {van Leeuwen}, {Walton}, {Arenou}, {Bastian}, {Cropper},
  {Drimmel}, {Katz}, {Lattanzi}, {Bakker}, {Cacciari}, {Casta{\~n}eda},
  {Chaoul}, {Cheek}, {De Angeli}, {Fabricius}, {Guerra}, {Holl}, {Masana},
  {Messineo}, {Mowlavi}, {Nienartowicz}, {Panuzzo}, {Portell}, {Riello},
  {Seabroke}, {Tanga}, {Th{\'e}venin}, {Gracia-Abril}, {Comoretto},
  {Garcia-Reinaldos}, {Teyssier}, {Altmann}, {Andrae}, {Audard},
  {Bellas-Velidis}, {Benson}, {Berthier}, {Blomme}, {Burgess}, {Busso},
  {Carry}, {Cellino}, {Clementini}, {Clotet}, {Creevey}, {Davidson}, {De
  Ridder}, {Delchambre}, {Dell'Oro}, {Ducourant},
  {Fern{\'a}ndez-Hern{\'a}ndez}, {Fouesneau}, {Fr{\'e}mat}, {Galluccio},
  {Garc{\'\i}a-Torres}, {Gonz{\'a}lez-N{\'u}{\~n}ez}, {Gonz{\'a}lez-Vidal},
  {Gosset}, {Guy}, {Halbwachs}, {Hambly}, {Harrison}, {Hern{\'a}ndez},
  {Hestroffer}, {Hodgkin}, {Hutton}, {Jasniewicz}, {Jean-Antoine-Piccolo},
  {Jordan}, {Korn}, {Krone-Martins}, {Lanzafame}, {Lebzelter}, {L{\"o}ffler},
  {Manteiga}, {Marrese}, {Mart{\'\i}n-Fleitas}, {Moitinho}, {Mora}, {Muinonen},
  {Osinde}, {Pancino}, {Pauwels}, {Petit}, {Recio-Blanco}, {Richards},
  {Rimoldini}, {Robin}, {Sarro}, {Siopis}, {Smith}, {Sozzetti}, {S{\"u}veges},
  {Torra}, {van Reeven}, {Abbas}, {Abreu Aramburu}, {Accart}, {Aerts},
  {Altavilla}, {{\'A}lvarez}, {Alvarez}, {Alves}, {Anderson}, {Andrei},
  {Anglada Varela}, {Antiche}, {Antoja}, {Arcay}, {Astraatmadja}, {Bach},
  {Baker}, {Balaguer-N{\'u}{\~n}ez}, {Balm}, {Barache}, {Barata}, {Barbato},
  {Barblan}, {Barklem}, {Barrado}, {Barros}, {Barstow}, {Bartholom{\'e}
  Mu{\~n}oz}, {Bassilana}, {Becciani}, {Bellazzini}, {Berihuete}, {Bertone},
  {Bianchi}, {Bienaym{\'e}}, {Blanco-Cuaresma}, {Boch}, {Boeche}, {Bombrun},
  {Borrachero}, {Bossini}, {Bouquillon}, {Bourda}, {Bragaglia}, {Bramante},
  {Breddels}, {Bressan}, {Brouillet}, {Br{\"u}semeister}, {Brugaletta},
  {Bucciarelli}, {Burlacu}, {Busonero}, {Butkevich}, {Buzzi}, {Caffau},
  {Cancelliere}, {Cannizzaro}, {Cantat-Gaudin}, {Carballo}, {Carlucci},
  {Carrasco}, {Casamiquela}, {Castellani}, {Castro-Ginard}, {Charlot},
  {Chemin}, {Chiavassa}, {Cocozza}, {Costigan}, {Cowell}, {Crifo}, {Crosta},
  {Crowley}, {Cuypers}, {Dafonte}, {Damerdji}, {Dapergolas}, {David}, {David},
  {de Laverny}, {De Luise}, {De March}, {de Martino}, {de Souza}, {de Torres},
  {Debosscher}, {del Pozo}, {Delbo}, {Delgado}, {Delgado}, {Di Matteo},
  {Diakite}, {Diener}, {Distefano}, {Dolding}, {Drazinos}, {Dur{\'a}n},
  {Edvardsson}, {Enke}, {Eriksson}, {Esquej}, {Eynard Bontemps}, {Fabre},
  {Fabrizio}, {Faigler}, {Falc{\~a}o}, {Farr{\`a}s Casas}, {Federici},
  {Fedorets}, {Fernique}, {Figueras}, {Filippi}, {Findeisen}, {Fonti},
  {Fraile}, {Fraser}, {Fr{\'e}zouls}, {Gai}, {Galleti}, {Garabato},
  {Garc{\'\i}a-Sedano}, {Garofalo}, {Garralda}, {Gavel}, {Gavras}, {Gerssen},
  {Geyer}, {Giacobbe}, {Gilmore}, {Girona}, {Giuffrida}, {Glass}, {Gomes},
  {Granvik}, {Gueguen}, {Guerrier}, {Guiraud}, {Guti{\'e}rrez-S{\'a}nchez},
  {Haigron}, {Hatzidimitriou}, {Hauser}, {Haywood}, {Heiter}, {Helmi}, {Heu},
  {Hilger}, {Hobbs}, {Hofmann}, {Holland}, {Huckle}, {Hypki}, {Icardi},
  {Jan{\ss}en}, {Jevardat de Fombelle}, {Jonker}, {Juh{\'a}sz}, {Julbe},
  {Karampelas}, {Kewley}, {Klar}, {Kochoska}, {Kohley}, {Kolenberg},
  {Kontizas}, {Kontizas}, {Koposov}, {Kordopatis}, {Kostrzewa-Rutkowska},
  {Koubsky}, {Lambert}, {Lanza}, {Lasne}, {Lavigne}, {Le Fustec}, {Le
  Poncin-Lafitte}, {Lebreton}, {Leccia}, {Leclerc}, {Lecoeur-Taibi},
  {Lenhardt}, {Leroux}, {Liao}, {Licata}, {Lindstr{\o}m}, {Lister}, {Livanou},
  {Lobel}, {L{\'o}pez}, {Managau}, {Mann}, {Mantelet}, {Marchal}, {Marchant},
  {Marconi}, {Marinoni}, {Marschalk{\'o}}, {Marshall}, {Martino}, {Marton},
  {Mary}, {Massari}, {Matijevi{\v{c}}}, {Mazeh}, {McMillan}, {Messina},
  {Michalik}, {Millar}, {Molina}, {Molinaro}, {Moln{\'a}r}, {Montegriffo},
  {Mor}, {Morbidelli}, {Morel}, {Morris}, {Mulone}, {Muraveva}, {Musella},
  {Nelemans}, {Nicastro}, {Noval}, {O'Mullane}, {Ord{\'e}novic},
  {Ord{\'o}{\~n}ez-Blanco}, {Osborne}, {Pagani}, {Pagano}, {Pailler},
  {Palacin}, {Palaversa}, {Panahi}, {Pawlak}, {Piersimoni}, {Pineau}, {Plachy},
  {Plum}, {Poggio}, {Poujoulet}, {Pr{\v{s}}a}, {Pulone}, {Racero}, {Ragaini},
  {Rambaux}, {Ramos-Lerate}, {Regibo}, {Reyl{\'e}}, {Riclet}, {Ripepi}, {Riva},
  {Rivard}, {Rixon}, {Roegiers}, {Roelens}, {Romero-G{\'o}mez}, {Rowell},
  {Royer}, {Ruiz-Dern}, {Sadowski}, {Sagrist{\`a} Sell{\'e}s}, {Sahlmann},
  {Salgado}, {Salguero}, {Sanna}, {Santana-Ros}, {Sarasso}, {Savietto},
  {Schultheis}, {Sciacca}, {Segol}, {Segovia}, {S{\'e}gransan}, {Shih},
  {Siltala}, {Silva}, {Smart}, {Smith}, {Solano}, {Solitro}, {Sordo}, {Soria
  Nieto}, {Souchay}, {Spagna}, {Spoto}, {Stampa}, {Steele},
  {Steidelm{\"u}ller}, {Stephenson}, {Stoev}, {Suess}, {Surdej}, {Szabados},
  {Szegedi-Elek}, {Tapiador}, {Taris}, {Tauran}, {Taylor}, {Teixeira},
  {Terrett}, {Teyssand ier}, {Thuillot}, {Titarenko}, {Torra Clotet}, {Turon},
  {Ulla}, {Utrilla}, {Uzzi}, {Vaillant}, {Valentini}, {Valette}, {van Elteren},
  {Van Hemelryck}, {van Leeuwen}, {Vaschetto}, {Vecchiato}, {Veljanoski},
  {Viala}, {Vicente}, {Vogt}, {von Essen}, {Voss}, {Votruba}, {Voutsinas},
  {Walmsley}, {Weiler}, {Wertz}, {Wevers}, {Wyrzykowski}, {Yoldas},
  {{\v{Z}}erjal}, {Ziaeepour}, {Zorec}, {Zschocke}, {Zucker}, {Zurbach}, \&
  {Zwitter}}]{DR22018}
{Gaia Collaboration}, {Brown}, A.~G.~A., {Vallenari}, A., {et~al.} 2018, \aap,
  616, A1, \dodoi{10.1051/0004-6361/201833051}

\bibitem[{{Galli} {et~al.}(2020){Galli}, {Bouy}, {Olivares}, {Miret-Roig},
  {Sarro}, {Barrado}, {Berihuete}, \& {Brandner}}]{Galli2020}
{Galli}, P.~A.~B., {Bouy}, H., {Olivares}, J., {et~al.} 2020, arXiv e-prints,
  arXiv:2001.05190.
\newblock \doarXiv{2001.05190}

\bibitem[{{Greaves} \& {Rice}(2010)}]{GR2010}
{Greaves}, J.~S., \& {Rice}, W.~K.~M. 2010, \mnras, 407, 1981,
  \dodoi{10.1111/j.1365-2966.2010.17043.x}

\bibitem[{{Grenier} {et~al.}(2015){Grenier}, {Black}, \& {Strong}}]{GBS2015}
{Grenier}, I.~A., {Black}, J.~H., \& {Strong}, A.~W. 2015, \araa, 53, 199,
  \dodoi{10.1146/annurev-astro-082214-122457}

\bibitem[{{Guarcello} {et~al.}(2016){Guarcello}, {Drake}, {Wright},
  {Albacete-Colombo}, {Clarke}, {Ercolano}, {Flaccomio}, {Kashyap}, {Micela},
  {Naylor}, {Schneider}, {Sciortino}, \& {Vink}}]{Guarcello2016}
{Guarcello}, M.~G., {Drake}, J.~J., {Wright}, N.~J., {et~al.} 2016, arXiv
  e-prints, arXiv:1605.01773.
\newblock \doarXiv{1605.01773}

\bibitem[{{Haworth} {et~al.}(2016){Haworth}, {Boubert}, {Facchini}, {Bisbas},
  \& {Clarke}}]{Haworth2016}
{Haworth}, T.~J., {Boubert}, D., {Facchini}, S., {Bisbas}, T.~G., \& {Clarke},
  C.~J. 2016, \mnras, 463, 3616, \dodoi{10.1093/mnras/stw2280}

\bibitem[{{Haworth} {et~al.}(2017){Haworth}, {Facchini}, {Clarke}, \&
  {Cleeves}}]{Haworth2017}
{Haworth}, T.~J., {Facchini}, S., {Clarke}, C.~J., \& {Cleeves}, L.~I. 2017,
  \mnras, 468, L108, \dodoi{10.1093/mnrasl/slx037}

\bibitem[{{Haworth} {et~al.}(2018){Haworth}, {Facchini}, {Clarke}, \&
  {Mohanty}}]{Haworth2018}
{Haworth}, T.~J., {Facchini}, S., {Clarke}, C.~J., \& {Mohanty}, S. 2018,
  \mnras, 475, 5460, \dodoi{10.1093/mnras/sty168}

\bibitem[{{Haworth} \& {Owen}(2020)}]{Haworth2020}
{Haworth}, T.~J., \& {Owen}, J.~E. 2020, \mnras, 492, 5030,
  \dodoi{10.1093/mnras/staa151}

\bibitem[{{Henney} \& {O'Dell}(1999)}]{Henney1999}
{Henney}, W.~J., \& {O'Dell}, C.~R. 1999, \aj, 118, 2350,
  \dodoi{10.1086/301087}

\bibitem[{{Herczeg} \& {Hillenbrand}(2015)}]{HH2015}
{Herczeg}, G.~J., \& {Hillenbrand}, L.~A. 2015, \apj, 808, 23,
  \dodoi{10.1088/0004-637X/808/1/23}

\bibitem[{{Hern{\'a}ndez} {et~al.}(2009){Hern{\'a}ndez}, {Calvet}, {Hartmann},
  {Muzerolle}, {Gutermuth}, \& {Stauffer}}]{Hernandez2009}
{Hern{\'a}ndez}, J., {Calvet}, N., {Hartmann}, L., {et~al.} 2009, \apj, 707,
  705, \dodoi{10.1088/0004-637X/707/1/705}

\bibitem[{{Hern{\'a}ndez} {et~al.}(2010){Hern{\'a}ndez}, {Morales-Calderon},
  {Calvet}, {Hartmann}, {Muzerolle}, {Gutermuth}, {Luhman}, \&
  {Stauffer}}]{Hernandez2010}
{Hern{\'a}ndez}, J., {Morales-Calderon}, M., {Calvet}, N., {et~al.} 2010, \apj,
  722, 1226, \dodoi{10.1088/0004-637X/722/2/1226}

\bibitem[{{Hern{\'a}ndez} {et~al.}(2007){Hern{\'a}ndez}, {Calvet},
  {Brice{\~n}o}, {Hartmann}, {Vivas}, {Muzerolle}, {Downes}, {Allen}, \&
  {Gutermuth}}]{Hernandez2007}
{Hern{\'a}ndez}, J., {Calvet}, N., {Brice{\~n}o}, C., {et~al.} 2007, \apj, 671,
  1784, \dodoi{10.1086/522882}

\bibitem[{{Hildebrand}(1983)}]{Hildebrand1983}
{Hildebrand}, R.~H. 1983, \qjras, 24, 267

\bibitem[{{Hollenbach} {et~al.}(1994){Hollenbach}, {Johnstone}, {Lizano}, \&
  {Shu}}]{Hollenbach1994}
{Hollenbach}, D., {Johnstone}, D., {Lizano}, S., \& {Shu}, F. 1994, \apj, 428,
  654, \dodoi{10.1086/174276}

\bibitem[{{Hoogerwerf} {et~al.}(2000){Hoogerwerf}, {de Bruijne}, \& {de
  Zeeuw}}]{Hoogerwerf2000}
{Hoogerwerf}, R., {de Bruijne}, J.~H.~J., \& {de Zeeuw}, P.~T. 2000, \apjl,
  544, L133, \dodoi{10.1086/317315}

\bibitem[{{Hoogerwerf} {et~al.}(2001){Hoogerwerf}, {de Bruijne}, \& {de
  Zeeuw}}]{Hoogerwerf2001}
---. 2001, \aap, 365, 49, \dodoi{10.1051/0004-6361:20000014}

\bibitem[{Hunter(2007)}]{Hunter2007}
Hunter, J.~D. 2007, Computing In Science \& Engineering, 9, 90,
  \dodoi{10.1109/MCSE.2007.55}

\bibitem[{{Indriolo} {et~al.}(2010){Indriolo}, {Blake}, {Goto}, {Usuda}, {Oka},
  {Geballe}, {Fields}, \& {McCall}}]{Indriolo2010}
{Indriolo}, N., {Blake}, G.~A., {Goto}, M., {et~al.} 2010, \apj, 724, 1357,
  \dodoi{10.1088/0004-637X/724/2/1357}

\bibitem[{{Johnstone} {et~al.}(1998){Johnstone}, {Hollenbach}, \&
  {Bally}}]{Johnstone1998}
{Johnstone}, D., {Hollenbach}, D., \& {Bally}, J. 1998, \apj, 499, 758,
  \dodoi{10.1086/305658}

\bibitem[{{Kama} {et~al.}(2015){Kama}, {Folsom}, \& {Pinilla}}]{Kama2015}
{Kama}, M., {Folsom}, C.~P., \& {Pinilla}, P. 2015, \aap, 582, L10,
  \dodoi{10.1051/0004-6361/201527094}

\bibitem[{{Kenyon} \& {Hartmann}(1995)}]{KH1995}
{Kenyon}, S.~J., \& {Hartmann}, L. 1995, \apjs, 101, 117,
  \dodoi{10.1086/192235}

\bibitem[{{Kim} {et~al.}(2016){Kim}, {Clarke}, {Fang}, \& {Facchini}}]{Kim2016}
{Kim}, J.~S., {Clarke}, C.~J., {Fang}, M., \& {Facchini}, S. 2016, \apjl, 826,
  L15, \dodoi{10.3847/2041-8205/826/1/L15}

\bibitem[{{Kounkel} {et~al.}(2018){Kounkel}, {Covey}, {Su{\'a}rez},
  {Rom{\'a}n-Z{\'u}{\~n}iga}, {Hernandez}, {Stassun}, {Jaehnig}, {Feigelson},
  {Pe{\~n}a Ram{\'\i}rez}, {Roman-Lopes}, {Da Rio}, {Stringfellow}, {Kim},
  {Borissova}, {Fern{\'a}ndez-Trincado}, {Burgasser},
  {Garc{\'\i}a-Hern{\'a}ndez}, {Zamora}, {Pan}, \& {Nitschelm}}]{Kounkel2018}
{Kounkel}, M., {Covey}, K., {Su{\'a}rez}, G., {et~al.} 2018, \aj, 156, 84,
  \dodoi{10.3847/1538-3881/aad1f1}

\bibitem[{{Kurtovic} {et~al.}(2018){Kurtovic}, {P{\'e}rez}, {Benisty}, {Zhu},
  {Zhang}, {Huang}, {Andrews}, {Dullemond}, {Isella}, {Bai}, {Carpenter},
  {Guzm{\'a}n}, {Ricci}, \& {Wilner}}]{Kurtovic2018}
{Kurtovic}, N.~T., {P{\'e}rez}, L.~M., {Benisty}, M., {et~al.} 2018, \apjl,
  869, L44, \dodoi{10.3847/2041-8213/aaf746}

\bibitem[{{Lang} {et~al.}(2000){Lang}, {Masheder}, {Dame}, \&
  {Thaddeus}}]{Lang2000}
{Lang}, W.~J., {Masheder}, M.~R.~W., {Dame}, T.~M., \& {Thaddeus}, P. 2000,
  \aap, 357, 1001

\bibitem[{{Lavalley} {et~al.}(1992){Lavalley}, {Isobe}, \&
  {Feigelson}}]{Lavalley1992}
{Lavalley}, M., {Isobe}, T., \& {Feigelson}, E. 1992, Astronomical Society of
  the Pacific Conference Series, Vol.~25, {ASURV: Astronomy Survival Analysis
  Package}, ed. D.~M. {Worrall}, C.~{Biemesderfer}, \& J.~{Barnes}, 245

\bibitem[{{Le Petit} {et~al.}(2016){Le Petit}, {Ruaud}, {Bron}, {Godard},
  {Roueff}, {Languignon}, \& {Le Bourlot}}]{LePetit2016}
{Le Petit}, F., {Ruaud}, M., {Bron}, E., {et~al.} 2016, \aap, 585, A105,
  \dodoi{10.1051/0004-6361/201526658}

\bibitem[{{Lee} {et~al.}(2015){Lee}, {Seon}, \& {Jo}}]{Lee2015}
{Lee}, D., {Seon}, K.-I., \& {Jo}, Y.-S. 2015, \apj, 806, 274,
  \dodoi{10.1088/0004-637X/806/2/274}

\bibitem[{{Lucas} {et~al.}(2020){Lucas}, {Bonnell}, \& {Dale}}]{Lucas2020}
{Lucas}, W.~E., {Bonnell}, I.~A., \& {Dale}, J.~E. 2020, \mnras, 493, 4700,
  \dodoi{10.1093/mnras/staa451}

\bibitem[{{Luhman} \& {Esplin}(2020)}]{LE2020}
{Luhman}, K.~L., \& {Esplin}, T.~L. 2020, \aj, 160, 44,
  \dodoi{10.3847/1538-3881/ab9599}

\bibitem[{{Lynden-Bell} \& {Pringle}(1974)}]{LBP1974}
{Lynden-Bell}, D., \& {Pringle}, J.~E. 1974, \mnras, 168, 603,
  \dodoi{10.1093/mnras/168.3.603}

\bibitem[{{Maddalena} \& {Morris}(1987)}]{MM1987}
{Maddalena}, R.~J., \& {Morris}, M. 1987, \apj, 323, 179,
  \dodoi{10.1086/165818}

\bibitem[{{Majewski} {et~al.}(2017){Majewski}, {Schiavon}, {Frinchaboy},
  {Allende Prieto}, {Barkhouser}, {Bizyaev}, {Blank}, {Brunner}, {Burton},
  {Carrera}, {Chojnowski}, {Cunha}, {Epstein}, {Fitzgerald}, {Garc{\'\i}a
  P{\'e}rez}, {Hearty}, {Henderson}, {Holtzman}, {Johnson}, {Lam}, {Lawler},
  {Maseman}, {M{\'e}sz{\'a}ros}, {Nelson}, {Nguyen}, {Nidever}, {Pinsonneault},
  {Shetrone}, {Smee}, {Smith}, {Stolberg}, {Skrutskie}, {Walker}, {Wilson},
  {Zasowski}, {Anders}, {Basu}, {Beland}, {Blanton}, {Bovy}, {Brownstein},
  {Carlberg}, {Chaplin}, {Chiappini}, {Eisenstein}, {Elsworth}, {Feuillet},
  {Fleming}, {Galbraith-Frew}, {Garc{\'\i}a}, {Garc{\'\i}a-Hern{\'a}ndez},
  {Gillespie}, {Girardi}, {Gunn}, {Hasselquist}, {Hayden}, {Hekker}, {Ivans},
  {Kinemuchi}, {Klaene}, {Mahadevan}, {Mathur}, {Mosser}, {Muna}, {Munn},
  {Nichol}, {O'Connell}, {Parejko}, {Robin}, {Rocha-Pinto}, {Schultheis},
  {Serenelli}, {Shane}, {Silva Aguirre}, {Sobeck}, {Thompson}, {Troup},
  {Weinberg}, \& {Zamora}}]{Majewski2017}
{Majewski}, S.~R., {Schiavon}, R.~P., {Frinchaboy}, P.~M., {et~al.} 2017, \aj,
  154, 94, \dodoi{10.3847/1538-3881/aa784d}

\bibitem[{{Manara} {et~al.}(2018{\natexlab{a}}){Manara}, {Morbidelli}, \&
  {Guillot}}]{Manara2018b}
{Manara}, C.~F., {Morbidelli}, A., \& {Guillot}, T. 2018{\natexlab{a}}, \aap,
  618, L3, \dodoi{10.1051/0004-6361/201834076}

\bibitem[{{Manara} {et~al.}(2018{\natexlab{b}}){Manara}, {Prusti}, {Comeron},
  {Mor}, {Alcal{\'a}}, {Antoja}, {Facchini}, {Fedele}, {Frasca}, {Jerabkova},
  {Rosotti}, {Spezzi}, \& {Spina}}]{Manara2018}
{Manara}, C.~F., {Prusti}, T., {Comeron}, F., {et~al.} 2018{\natexlab{b}},
  \aap, 615, L1, \dodoi{10.1051/0004-6361/201833383}

\bibitem[{{Manara} {et~al.}(2020){Manara}, {Natta}, {Rosotti}, {Alcala},
  {Nisini}, {Lodato}, {Testi}, {Pascucci}, {Hillenbrand}, {Carpenter},
  {Scholz}, {Fedele}, {Frasca}, {Mulders}, {Rigliaco}, {Scardoni}, \&
  {Zari}}]{Manara2020}
{Manara}, C.~F., {Natta}, A., {Rosotti}, G.~P., {et~al.} 2020, arXiv e-prints,
  arXiv:2004.14232.
\newblock \doarXiv{2004.14232}

\bibitem[{{Mann} {et~al.}(2014){Mann}, {Di Francesco}, {Johnstone}, {Andrews},
  {Williams}, {Bally}, {Ricci}, {Hughes}, \& {Matthews}}]{Mann2014}
{Mann}, R.~K., {Di Francesco}, J., {Johnstone}, D., {et~al.} 2014, \apj, 784,
  82, \dodoi{10.1088/0004-637X/784/1/82}

\bibitem[{{Mathews} {et~al.}(2012){Mathews}, {Williams}, \&
  {M{\'e}nard}}]{Mathews2012}
{Mathews}, G.~S., {Williams}, J.~P., \& {M{\'e}nard}, F. 2012, \apj, 753, 59,
  \dodoi{10.1088/0004-637X/753/1/59}

\bibitem[{Mathieu(2015)}]{Mathieu2015}
Mathieu, R. 2015, $\lambda$ Ori: A Case Study in Star Formation, 147--163,
  \dodoi{10.1007/978-3-662-47290-3_11}

\bibitem[{{Mathieu}(2008)}]{Mathieu2008}
{Mathieu}, R.~D. 2008, {The {\ensuremath{\lambda}} Orionis Star Forming
  Region}, ed. B.~{Reipurth}, Vol.~4, 757

\bibitem[{{Maxted} {et~al.}(2008){Maxted}, {Jeffries}, {Oliveira}, {Naylor}, \&
  {Jackson}}]{Maxted2008}
{Maxted}, P.~F.~L., {Jeffries}, R.~D., {Oliveira}, J.~M., {Naylor}, T., \&
  {Jackson}, R.~J. 2008, \mnras, 385, 2210,
  \dodoi{10.1111/j.1365-2966.2008.13008.x}

\bibitem[{{McMullin} {et~al.}(2007){McMullin}, {Waters}, {Schiebel}, {Young},
  \& {Golap}}]{CASA2007}
{McMullin}, J.~P., {Waters}, B., {Schiebel}, D., {Young}, W., \& {Golap}, K.
  2007, Astronomical Society of the Pacific Conference Series, Vol. 376, {CASA
  Architecture and Applications}, ed. R.~A. {Shaw}, F.~{Hill}, \& D.~J. {Bell},
  127

\bibitem[{{Miotello} {et~al.}(2017){Miotello}, {van Dishoeck}, {Williams},
  {Ansdell}, {Guidi}, {Hogerheijde}, {Manara}, {Tazzari}, {Testi}, {van der
  Marel}, \& {van Terwisga}}]{Miotello2017}
{Miotello}, A., {van Dishoeck}, E.~F., {Williams}, J.~P., {et~al.} 2017, \aap,
  599, A113, \dodoi{10.1051/0004-6361/201629556}

\bibitem[{{Miotello} {et~al.}(2019){Miotello}, {Facchini}, {van Dishoeck},
  {Cazzoletti}, {Testi}, {Williams}, {Ansdell}, {van Terwisga}, \& {van der
  Marel}}]{Miotello2019}
{Miotello}, A., {Facchini}, S., {van Dishoeck}, E.~F., {et~al.} 2019, \aap,
  631, A69, \dodoi{10.1051/0004-6361/201935441}

\bibitem[{{Muley} {et~al.}(2019){Muley}, {Fung}, \& {van der
  Marel}}]{Muley2019}
{Muley}, D., {Fung}, J., \& {van der Marel}, N. 2019, \apjl, 879, L2,
  \dodoi{10.3847/2041-8213/ab24d0}

\bibitem[{{Murdin} \& {Penston}(1977)}]{MP1977}
{Murdin}, P., \& {Penston}, M.~V. 1977, \mnras, 181, 657,
  \dodoi{10.1093/mnras/181.4.657}

\bibitem[{{Najita} \& {Kenyon}(2014)}]{NK2014}
{Najita}, J.~R., \& {Kenyon}, S.~J. 2014, \mnras, 445, 3315,
  \dodoi{10.1093/mnras/stu1994}

\bibitem[{{Nielsen} {et~al.}(2019){Nielsen}, {De Rosa}, {Macintosh}, {Wang},
  {Ruffio}, {Chiang}, {Marley}, {Saumon}, {Savransky}, {Ammons}, {Bailey},
  {Barman}, {Blain}, {Bulger}, {Burrows}, {Chilcote}, {Cotten}, {Czekala},
  {Doyon}, {Duch{\^e}ne}, {Esposito}, {Fabrycky}, {Fitzgerald}, {Follette},
  {Fortney}, {Gerard}, {Goodsell}, {Graham}, {Greenbaum}, {Hibon}, {Hinkley},
  {Hirsch}, {Hom}, {Hung}, {Dawson}, {Ingraham}, {Kalas}, {Konopacky},
  {Larkin}, {Lee}, {Lin}, {Maire}, {Marchis}, {Marois}, {Metchev},
  {Millar-Blanchaer}, {Morzinski}, {Oppenheimer}, {Palmer}, {Patience},
  {Perrin}, {Poyneer}, {Pueyo}, {Rafikov}, {Rajan}, {Rameau}, {Rantakyr{\"o}},
  {Ren}, {Schneider}, {Sivaramakrishnan}, {Song}, {Soummer}, {Tallis},
  {Thomas}, {Ward-Duong}, \& {Wolff}}]{Nielsen2019}
{Nielsen}, E.~L., {De Rosa}, R.~J., {Macintosh}, B., {et~al.} 2019, \aj, 158,
  13, \dodoi{10.3847/1538-3881/ab16e9}

\bibitem[{{Pascucci} {et~al.}(2016){Pascucci}, {Testi}, {Herczeg}, {Long},
  {Manara}, {Hendler}, {Mulders}, {Krijt}, {Ciesla}, {Henning}, {Mohanty},
  {Drabek-Maunder}, {Apai}, {Sz{\H{u}}cs}, {Sacco}, \&
  {Olofsson}}]{Pascucci2016}
{Pascucci}, I., {Testi}, L., {Herczeg}, G.~J., {et~al.} 2016, \apj, 831, 125,
  \dodoi{10.3847/0004-637X/831/2/125}

\bibitem[{{P{\'e}rez} {et~al.}(2018){P{\'e}rez}, {Benisty}, {Andrews},
  {Isella}, {Dullemond}, {Huang}, {Kurtovic}, {Guzm{\'a}n}, {Zhu}, {Birnstiel},
  {Zhang}, {Carpenter}, {Wilner}, {Ricci}, {Bai}, {Weaver}, \&
  {{\"O}berg}}]{Perez2018}
{P{\'e}rez}, L.~M., {Benisty}, M., {Andrews}, S.~M., {et~al.} 2018, \apjl, 869,
  L50, \dodoi{10.3847/2041-8213/aaf745}

\bibitem[{{Pinilla} {et~al.}(2020){Pinilla}, {Pascucci}, \&
  {Marino}}]{Pinilla2020}
{Pinilla}, P., {Pascucci}, I., \& {Marino}, S. 2020, \aap, 635, A105,
  \dodoi{10.1051/0004-6361/201937003}

\bibitem[{{Pinilla} {et~al.}(2018){Pinilla}, {Benisty}, {de Boer}, {Manara},
  {Bouvier}, {Dominik}, {Ginski}, {Loomis}, \& {Sicilia Aguilar}}]{Pinilla2018}
{Pinilla}, P., {Benisty}, M., {de Boer}, J., {et~al.} 2018, \apj, 868, 85,
  \dodoi{10.3847/1538-4357/aae824}

\bibitem[{{Ribas} {et~al.}(2015){Ribas}, {Bouy}, \& {Mer{\'\i}n}}]{Ribas2015}
{Ribas}, {\'A}., {Bouy}, H., \& {Mer{\'\i}n}, B. 2015, \aap, 576, A52,
  \dodoi{10.1051/0004-6361/201424846}

\bibitem[{{Rieke} {et~al.}(2004){Rieke}, {Young}, {Engelbracht}, {Kelly},
  {Low}, {Haller}, {Beeman}, {Gordon}, {Stansberry}, {Misselt}, {Cadien},
  {Morrison}, {Rivlis}, {Latter}, {Noriega-Crespo}, {Padgett}, {Stapelfeldt},
  {Hines}, {Egami}, {Muzerolle}, {Alonso-Herrero}, {Blaylock}, {Dole}, {Hinz},
  {Le Floc'h}, {Papovich}, {P{\'e}rez-Gonz{\'a}lez}, {Smith}, {Su}, {Bennett},
  {Frayer}, {Henderson}, {Lu}, {Masci}, {Pesenson}, {Rebull}, {Rho}, {Keene},
  {Stolovy}, {Wachter}, {Wheaton}, {Werner}, \& {Richards}}]{Rieke2004}
{Rieke}, G.~H., {Young}, E.~T., {Engelbracht}, C.~W., {et~al.} 2004, \apjs,
  154, 25, \dodoi{10.1086/422717}

\bibitem[{{Rigliaco} {et~al.}(2009){Rigliaco}, {Natta}, {Randich}, \&
  {Sacco}}]{Rigliaco2009}
{Rigliaco}, E., {Natta}, A., {Randich}, S., \& {Sacco}, G. 2009, \aap, 495,
  L13, \dodoi{10.1051/0004-6361/200811535}

\bibitem[{{Rosenfeld} {et~al.}(2012){Rosenfeld}, {Qi}, {Andrews}, {Wilner},
  {Corder}, {Dullemond}, {Lin}, {Hughes}, {D'Alessio}, \& {Ho}}]{Rosenfeld2012}
{Rosenfeld}, K.~A., {Qi}, C., {Andrews}, S.~M., {et~al.} 2012, \apj, 757, 129,
  \dodoi{10.1088/0004-637X/757/2/129}

\bibitem[{{Rosenthal} {et~al.}(2020){Rosenthal}, {Chiang}, {Ginzburg}, \&
  {Murray-Clay}}]{Rosenthal2020}
{Rosenthal}, M.~M., {Chiang}, E.~I., {Ginzburg}, S., \& {Murray-Clay}, R.~A.
  2020, \mnras, \dodoi{10.1093/mnras/staa1721}

\bibitem[{{Sacco} {et~al.}(2008){Sacco}, {Franciosini}, {Randich}, \&
  {Pallavicini}}]{Sacco2008}
{Sacco}, G.~G., {Franciosini}, E., {Randich}, S., \& {Pallavicini}, R. 2008,
  \aap, 488, 167, \dodoi{10.1051/0004-6361:20079049}

\bibitem[{{Scally} \& {Clarke}(2001)}]{SC2001}
{Scally}, A., \& {Clarke}, C. 2001, \mnras, 325, 449,
  \dodoi{10.1046/j.1365-8711.2001.04274.x}

\bibitem[{{Schwarz} {et~al.}(2018{\natexlab{a}}){Schwarz}, {Bergin}, {Cleeves},
  {Zhang}, {{\"O}berg}, {Blake}, \& {Anderson}}]{Scharz2018}
{Schwarz}, K.~R., {Bergin}, E.~A., {Cleeves}, L.~I., {et~al.}
  2018{\natexlab{a}}, \apj, 856, 85, \dodoi{10.3847/1538-4357/aaae08}

\bibitem[{{Schwarz} {et~al.}(2018{\natexlab{b}}){Schwarz}, {Bergin}, {Cleeves},
  {Zhang}, {{\"O}berg}, {Blake}, \& {Anderson}}]{Schwarz2018}
---. 2018{\natexlab{b}}, \apj, 856, 85, \dodoi{10.3847/1538-4357/aaae08}

\bibitem[{{Sellek} {et~al.}(2020){Sellek}, {Booth}, \& {Clarke}}]{Sellek2020}
{Sellek}, A.~D., {Booth}, R.~A., \& {Clarke}, C.~J. 2020, \mnras, 492, 1279,
  \dodoi{10.1093/mnras/stz3528}

\bibitem[{{Siess} {et~al.}(2000){Siess}, {Dufour}, \& {Forestini}}]{Siess2000}
{Siess}, L., {Dufour}, E., \& {Forestini}, M. 2000, \aap, 358, 593.
\newblock \doarXiv{astro-ph/0003477}

\bibitem[{{Skrutskie} {et~al.}(2006){Skrutskie}, {Cutri}, {Stiening},
  {Weinberg}, {Schneider}, {Carpenter}, {Beichman}, {Capps}, {Chester},
  {Elias}, {Huchra}, {Liebert}, {Lonsdale}, {Monet}, {Price}, {Seitzer},
  {Jarrett}, {Kirkpatrick}, {Gizis}, {Howard}, {Evans}, {Fowler}, {Fullmer},
  {Hurt}, {Light}, {Kopan}, {Marsh}, {McCallon}, {Tam}, {Van Dyk}, \&
  {Wheelock}}]{TWOMASS2006AJ}
{Skrutskie}, M.~F., {Cutri}, R.~M., {Stiening}, R., {et~al.} 2006, \aj, 131,
  1163, \dodoi{10.1086/498708}

\bibitem[{{St{\"o}rzer} \& {Hollenbach}(1999)}]{SH1999}
{St{\"o}rzer}, H., \& {Hollenbach}, D. 1999, \apj, 515, 669,
  \dodoi{10.1086/307055}

\bibitem[{{Takami} {et~al.}(2014){Takami}, {Hasegawa}, {Muto}, {Gu}, {Dong},
  {Karr}, {Hashimoto}, {Kusakabe}, {Chapillon}, {Tang}, {Itoh}, {Carson},
  {Follette}, {Mayama}, {Sitko}, {Janson}, {Grady}, {Kudo}, {Akiyama}, {Kwon},
  {Takahashi}, {Suenaga}, {Abe}, {Brandner}, {Brand t}, {Currie}, {Egner},
  {Feldt}, {Guyon}, {Hayano}, {Hayashi}, {Hayashi}, {Henning}, {Hodapp},
  {Honda}, {Ishii}, {Iye}, {Kandori}, {Knapp}, {Kuzuhara}, {McElwain},
  {Matsuo}, {Miyama}, {Morino}, {Moro-Martin}, {Nishimura}, {Pyo}, {Serabyn},
  {Suto}, {Suzuki}, {Takato}, {Terada}, {Thalmann}, {Tomono}, {Turner},
  {Wisniewski}, {Watanabe}, {Yamada}, {Takami}, {Usuda}, \&
  {Tamura}}]{Takami2014}
{Takami}, M., {Hasegawa}, Y., {Muto}, T., {et~al.} 2014, \apj, 795, 71,
  \dodoi{10.1088/0004-637X/795/1/71}

\bibitem[{{Tazzari} {et~al.}(2017){Tazzari}, {Testi}, {Natta}, {Ansdell},
  {Carpenter}, {Guidi}, {Hogerheijde}, {Manara}, {Miotello}, {van der Marel},
  {van Dishoeck}, \& {Williams}}]{Tazzari2017}
{Tazzari}, M., {Testi}, L., {Natta}, A., {et~al.} 2017, \aap, 606, A88,
  \dodoi{10.1051/0004-6361/201730890}

\bibitem[{{Trapman} {et~al.}(2020){Trapman}, {Ansdell}, {Hogerheijde},
  {Facchini}, {Manara}, {Miotello}, {Williams}, \& {Bruderer}}]{Trapman2020}
{Trapman}, L., {Ansdell}, M., {Hogerheijde}, M.~R., {et~al.} 2020, arXiv
  e-prints, arXiv:2004.07257.
\newblock \doarXiv{2004.07257}

\bibitem[{{Tripathi} {et~al.}(2017){Tripathi}, {Andrews}, {Birnstiel}, \&
  {Wilner}}]{Tripathi2017}
{Tripathi}, A., {Andrews}, S.~M., {Birnstiel}, T., \& {Wilner}, D.~J. 2017,
  \apj, 845, 44, \dodoi{10.3847/1538-4357/aa7c62}

\bibitem[{{van der Marel} {et~al.}(2018){van der Marel}, {Williams}, {Ansdell},
  {Manara}, {Miotello}, {Tazzari}, {Testi}, {Hogerheijde}, {Bruderer}, {van
  Terwisga}, \& {van Dishoeck}}]{vdMarel2018}
{van der Marel}, N., {Williams}, J.~P., {Ansdell}, M., {et~al.} 2018, \apj,
  854, 177, \dodoi{10.3847/1538-4357/aaaa6b}

\bibitem[{{van Terwisga} {et~al.}(2019{\natexlab{a}}){van Terwisga}, {Hacar},
  \& {van Dishoeck}}]{Terwisga2019}
{van Terwisga}, S.~E., {Hacar}, A., \& {van Dishoeck}, E.~F.
  2019{\natexlab{a}}, \aap, 628, A85, \dodoi{10.1051/0004-6361/201935378}

\bibitem[{{van Terwisga} {et~al.}(2019{\natexlab{b}}){van Terwisga}, {van
  Dishoeck}, {Cazzoletti}, {Facchini}, {Trapman}, {Williams}, {Manara},
  {Miotello}, {van der Marel}, {Ansdell}, {Hogerheijde}, {Tazzari}, \&
  {Testi}}]{vanTerwisga2019}
{van Terwisga}, S.~E., {van Dishoeck}, E.~F., {Cazzoletti}, P., {et~al.}
  2019{\natexlab{b}}, \aap, 623, A150, \dodoi{10.1051/0004-6361/201834257}

\bibitem[{{Williams}(2012)}]{Willialms2012}
{Williams}, J.~P. 2012, Meteoritics and Planetary Science, 47, 1915,
  \dodoi{10.1111/maps.12028}

\bibitem[{{Winn} \& {Fabrycky}(2015)}]{WF2015}
{Winn}, J.~N., \& {Fabrycky}, D.~C. 2015, \araa, 53, 409,
  \dodoi{10.1146/annurev-astro-082214-122246}

\bibitem[{{Winter} {et~al.}(2018){Winter}, {Clarke}, {Rosotti}, {Ih},
  {Facchini}, \& {Haworth}}]{Winter2018}
{Winter}, A.~J., {Clarke}, C.~J., {Rosotti}, G., {et~al.} 2018, \mnras, 478,
  2700, \dodoi{10.1093/mnras/sty984}

\bibitem[{{Winter} {et~al.}(2019){Winter}, {Clarke}, \& {Rosotti}}]{Winter2019}
{Winter}, A.~J., {Clarke}, C.~J., \& {Rosotti}, G.~P. 2019, \mnras, 485, 1489,
  \dodoi{10.1093/mnras/stz473}

\bibitem[{{Winter} {et~al.}(2020){Winter}, {Kruijssen}, {Chevance}, {Keller},
  \& {Longmore}}]{Winter2020}
{Winter}, A.~J., {Kruijssen}, J.~M.~D., {Chevance}, M., {Keller}, B.~W., \&
  {Longmore}, S.~N. 2020, \mnras, 491, 903, \dodoi{10.1093/mnras/stz2747}

\bibitem[{{Wyatt}(2008)}]{Wyatt2008}
{Wyatt}, M.~C. 2008, \araa, 46, 339,
  \dodoi{10.1146/annurev.astro.45.051806.110525}

\bibitem[{{Zamora-Avil{\'e}s} {et~al.}(2019){Zamora-Avil{\'e}s},
  {Ballesteros-Paredes}, {Hern{\'a}ndez}, {Rom{\'a}n-Z{\'u}{\~n}iga}, {Lora},
  \& {Kounkel}}]{Zamora2019}
{Zamora-Avil{\'e}s}, M., {Ballesteros-Paredes}, J., {Hern{\'a}ndez}, J.,
  {et~al.} 2019, \mnras, 488, 3406, \dodoi{10.1093/mnras/stz1897}

\bibitem[{{Zhu} {et~al.}(2011){Zhu}, {Nelson}, {Hartmann}, {Espaillat}, \&
  {Calvet}}]{Zhu2011}
{Zhu}, Z., {Nelson}, R.~P., {Hartmann}, L., {Espaillat}, C., \& {Calvet}, N.
  2011, \apj, 729, 47, \dodoi{10.1088/0004-637X/729/1/47}

\bibitem[{{Zhu} {et~al.}(2019){Zhu}, {Zhang}, {Jiang}, {Kataoka}, {Birnstiel},
  {Dullemond}, {Andrews}, {Huang}, {P{\'e}rez}, {Carpenter}, {Bai}, {Wilner},
  \& {Ricci}}]{Zhu2019}
{Zhu}, Z., {Zhang}, S., {Jiang}, Y.-F., {et~al.} 2019, \apjl, 877, L18,
  \dodoi{10.3847/2041-8213/ab1f8c}

\end{thebibliography}
\bibliographystyle{aasjournal}



\end{document}